\documentclass[preprint,prd,showpacs,showkeys,eqsecnum,nofootinbib,unsortedaddress]{revtex4}

\usepackage{amsfonts}
\usepackage{amsmath}
\usepackage{amssymb}
\usepackage{epsfig}
\input{tcilatex}

\begin{document}

\title{NLO Dispersion Laws for Slow-Moving Quarks in HTL QCD}
\author{Abdessamad Abada$^{\mathrm{a,b}}$}
\email{a.abada@uaeu.ac.ae}
\author{Karima Benchallal$^{\mathrm{a}}$}
\email{benchallal@ens-kouba.dz}
\author{Karima Bouakaz$^{\mathrm{a}}$}
\email{bouakaz@ens-kouba.dz}
\affiliation{$^{\mathrm{a}}$Laboratoire de Physique des Particules et Physique
Statistique, Ecole Normale Sup\'{e}rieure, BP 92 Vieux Kouba, 16050 Alger,
Algeria\\
$^{\mathrm{b}}$Physics Department, United Arab Emirates University, POB
17551 Al Ain, UAE}
\keywords{quark dispersion laws. QCD hard thermal loops. NLO.}
\pacs{11.10.Wx 12.38.Bx 12.38.Cy 12.38.Mh}

\begin{abstract}
We determine the next-to-leading order dispersion laws for slow-moving
quarks in hard-thermal-loop perturbation of high-temperature QCD where weak
coupling is assumed. Real-time formalism is used. The next-to-leading order
quark self-energy is written in terms of three and four HTL-dressed vertex
functions. The hard thermal loops contributing to these vertex functions are
calculated ab initio and expressed using the Feynman parametrization which
allows the calculation of the solid-angle integrals involved. We use a
prototype of the resulting integrals to indicate how finite results are
obtained in the limit of vanishing regularizer.
\end{abstract}

\date{\today }
\maketitle

\section{Introduction}

The past ten years or so have witnessed an abundant activity that tries to
understand the properties of the quark gluon plasma, the mechanism(s) of
deconfinement and the characteristics of the transition from hadronic matter
to quarks and gluons. Experimentally, collaborations at RHIC \cite
{rhic-collabos} and now ALICE at the LHC \cite{alice} are among the main
efforts dedicated to this aim. On the other hand, lattice simulations \cite
{lqcd} as well as hydrodynamic modeling \cite{hydro} help investigate the
thermodynamic and transport properties of the plasma.

From a perturbative QCD standpoint, calculations at high temperature use the
so-called hard-thermal-loop (HTL) summation of Feynman diagrams \cite
{HTL-perturbation}. For example, one determines the pressure and the quark
number susceptibilities from the thermodynamic potential calculated to two
and three-loop order \cite{two-three-loop-htl}, or the electric and magnetic
properties of the plasma \cite{liu-luo-wang-xu}.

HTL summation came about in order to overcome early problems encountered in
the standard loop-expansion of high-temperature QCD \cite{early-htl}.
However, it makes the next-to-leading order (NLO) dispersion relations for
slow-moving quasiparticles, quarks and gluons, difficult to calculate as
they involve the use of the fully HTL-dressed propagators and vertices. The
first quantity calculated in NLO fully-HTL-dressed perturbation is the
non-moving gluon damping rate \cite{gamt0}. That was followed by the
calculation of the non-moving quark damping rate\footnote{
This quantity was later calculated in \cite{carrington--PRD75-2007-045019}
using the real-time formalism.} \cite{gamq0}. These calculations have been
performed in the imaginary-time formalism of finite-temperature quantum
field theory \cite{imaginary-time-formalism}; they extract the damping rates
from the imaginary part of the fully-HTL-dressed one-loop order
self-energies after analytic continuation to real energies is taken. In a
line of works, we have used this formalism and looked into the infrared
behavior of fully-HTL-dressed one-loop-order damping rates of slow-moving
longitudinal \cite{gaml} and transverse gluons \cite{gamt}, quarks \cite
{gamq}, fermions\footnote{
Note that the fermion damping rate at zero momentum in finite-temperature
QED is independently calculated in \cite{carrington--PRD75-2007-045019}
using the real-time formalism. The same result is found.} \cite{fermions}
and photons \cite{photons} in QED, and quasiparticles \cite{sqed} in scalar
QED\footnote{
In scalar QED, we have also calculated the NLO energy of the quasiparticle.}.

In this logic, the natural step forward is to try to calculate the NLO
energies of the quasiparticles. That would come from the real part of the
fully-HTL-dressed one-loop order self-energies. This is notoriously much
harder than extracting the imaginary part. The first contribution in this
direction is the determination of the pure-gluon plasma frequency $\omega
_{g}\left( 0\right) $ at next-to-leading order in the long wavelength limit 
\cite{schulz}. Imaginary-time formalism is used and the number $N_{f}$ of
quark flavors is set to zero from the outset. A gauge-invariant result is
found: 
\begin{equation}
\omega _{g}\left( 0\right) =\frac{\sqrt{N_{c}}}{3}gT\left( 1-0.09\sqrt{N_{c}}
g+\dots \right) ,\qquad N_{f}=0.  \label{omegal0}
\end{equation}
In this result, $g$ is the strong coupling constant, $T$ the temperature and 
$N_{c}$ the number of colors. The next contribution came some time later 
\cite{carrington-et-al--EPJC50-2007-711}, \cite
{carrington--PRD75-2007-045019} and \cite
{carrington-et-al--PRD78-2008-045018}, namely the determination of the NLO
fermion mass $\omega _{q}\left( 0\right) $ in high-$T$ QCD (and QED). For
quarks, the result found is \cite{carrington-et-al--PRD78-2008-045018}: 
\begin{equation}
\omega _{q}\left( 0\right) =\frac{gT}{\sqrt{6}}\left( 1+\frac{1.87}{4\pi }
g+\dots \right) ,\qquad N_{c}=3,N_{f}=2.  \label{omegalq0}
\end{equation}
This calculation was performed in the real-time formalism of quantum field
theory (for reviews on this formalism, see \cite{real-time-formalism}).

One should note in this respect that the NLO contributions to such
quantities come for soft one-loop diagrams. Indeed, a general power-counting
analysis performed in \cite{mirza-carrington--PRD87-2013-065008} using the
real-time formalism shows that, except for the photon self-energy where
two-loop diagrams with hard internal momenta do contribute, next-to-leading
order contributions come from soft one-loop diagrams with HTL-dressed
vertices and propagators. The work \cite{carrington--PRD75-2007-045019}
shows that the usual power-counting in imaginary-time formalism
overestimates a number of terms that are in effect subleading.

The present work aims at determining the NLO dispersion relations, real part
(energy) and imaginary part (damping rate), for slow-moving quasi-quarks in
a quark-gluon plasma at high temperature with bare masses taken to zero. We
use the closed-time-path formulation of the real-time formalism of
finite-temperature quantum field theory \cite{martin-schwinger,keldysh}. The
advantage is that we avoid the analytic continuation from discrete Matsubara
frequencies to continuous real energies and all that comes with it which, in
a sophisticated calculation like the determination of the dispersion
relations, can make it difficult to extract the analytic behavior of the
physical quantities. But as everything comes with a price, one disadvantage
is that, as a result of the so-called doubling of degrees of freedom, each $
n $-point function acquires a tensor structure with $2^{n}$ components to
start with, which means a significant increase in the number of say one-loop
diagrams involving three and four-point 1PI vertex functions. In addition,
this calculation will not benefit from nice simplifications that arise when
we set the quark momentum to zero, like the replacement of momentum
contractions of HTL vertices with appropriate HTL\ self-energy differences
via Ward identities \cite{carrington-et-al--PRD78-2008-045018}.

This article is organized as follows. After this introduction, we define in
section two the HTL-dressed quark and gluon propagators, as well as the
quark energies and damping rates at next-to-leading order $g^{2}T$. These
quantities are directly related to the NLO fully-HTL-dressed quark
self-energy $\Sigma ^{(1)}$. This quantity is calculated in section three.
We give there an explicit expression of $\Sigma ^{(1)}$ in terms of the
three and four HTL-dressed vertex functions. These functions are derived ab
initio as discrepancies between different results in the literature are
found \cite{defu-et-al--PRD61-2000-085013,fueki-nakkagawa-yokota-yoshida}.
Then, in section four, we introduce a Feynman parametrization to help
perform the solid-angle integrals present in the vertex hard thermal loops.

Still, the subsequent integration task remains formidable. In section five,
we take a prototype and show how one can carry out with such integrals. The
work is mainly numerical. We choose to avoid using the spectral
decompositions of the HTL-dressed propagators and aim at getting a finite
result with the multi-integral as defined. We indicate in this section how
it is possible to obtain a stable behavior down to $10^{-8}$ in unit of the
quark thermal mass $m_{f}$.

Brief concluding remarks populate section six. An appendix is dedicated to
the derivation of the three and four-vertex hard thermal loops.

\section{The NLO dispersion relations}

We consider QCD with $N_{c}$ colors and $N_{f}$ flavors. The quark
dispersion relations can been cast as: 
\begin{equation}
\det \left( P\hspace{-8pt}/-\Sigma \left( P\right) \right) =0.
\label{fermion-dispersion-relation}
\end{equation}
Here, $P=\left( p_{0},\vec{p}\right) $ is the quark external soft
four-momentum and $\Sigma \left( P\right) $ is the quark self-energy, which
can be decomposed into two components $\Sigma _{\pm }$ in the following
manner: 
\begin{equation}
\Sigma \left( P\right) =\gamma _{+p}\Sigma _{-}\left( P\right) +\gamma
_{-p}\Sigma _{+}\left( P\right) \text{.}  \label{fermion-self-energy}
\end{equation}
In this expression, $\gamma _{\pm p}\equiv \left( \gamma ^{0}\mp \vec{\gamma}
. \hat{p}\right) /2$, with $\hat{p}=\vec{p}/p$ and $\left\{ \gamma ^{\mu
}\right\} $ the four Dirac matrices. Relation (\ref
{fermion-dispersion-relation}) is equivalent to the following two dispersion
relations: 
\begin{equation}
p_{0}\mp p-\Sigma _{\pm }\left( P\right) =0.
\label{helicity-projected-dispersions}
\end{equation}
On shell, the (complex) quark energy $p_{0}\equiv \Omega \left( p\right) $
can be decomposed in powers of the coupling constant $g$: 
\begin{equation}
\Omega \left( p\right) =\Omega ^{(0)}(p)+\Omega ^{(1)}(p)+\dots \,.
\label{fermion-energy-decomposition}
\end{equation}
This follows a similar decomposition of $\Sigma $, namely: 
\begin{equation}
\Sigma \left( P\right) =\Sigma _{\mathrm{HTL}}\left( P\right) +\Sigma
^{(1)}\left( P\right) +\dots ,  \label{Sigma-decomposition}
\end{equation}
where $\Sigma _{\mathrm{HTL}}$ is the lowest-order contribution, formed by
the hard thermal loops of order $gT$, and $\Sigma ^{(1)}$ the NLO
contribution, of order $g^{2}T$. The contribution $\Omega ^{\left( 0\right)
}\left( p\right) \equiv \omega _{\pm }\left( p\right) $ is thus of lowest
order $gT$, and $\Omega ^{(1)}(p)$ is the NLO contribution of order $g^{2}T$
. The dispersion relations (\ref{helicity-projected-dispersions}) can
therefore be decomposed as: 
\begin{equation}
\omega _{\pm }(p)+\Omega _{\pm }^{(1)}(p)+\dots =\pm p+\Sigma _{\mathrm{HTL}
\pm }\left( \Omega _{\pm }\left( p\right) ,p\right) +\Sigma _{\pm
}^{(1)}\left( \Omega _{\pm }\left( p\right) ,p\right) +\dots \,.
\label{dispersions-in-powers-g}
\end{equation}
Remembering that $p\sim gT$, We have: 
\begin{equation}
\Omega _{\pm }^{(1)}(p)=\frac{\Sigma _{\pm }^{(1)}\left( \omega _{\pm
}(p),p\right) }{1-\partial _{\omega }\left. \Sigma _{\mathrm{HTL}\pm }\left(
\omega ,p\right) \right\vert _{\omega =\omega _{\pm }(p)}}.
\label{NLO-complex-energy}
\end{equation}
Here $\partial _{\omega }$ stands for $\partial /\partial \omega $. The real
parts of $\Omega _{\pm }^{(1)}(p)$ are the NLO corrections $\omega _{\pm
}^{(1)}\left( p\right) $ to the plasma quark energies, and the negatives of
the imaginary parts are their damping rates $\gamma _{\pm }\left( p\right) $.

\begin{figure}[htb]
\centering
\includegraphics[width=6.1in,height=3.1in]{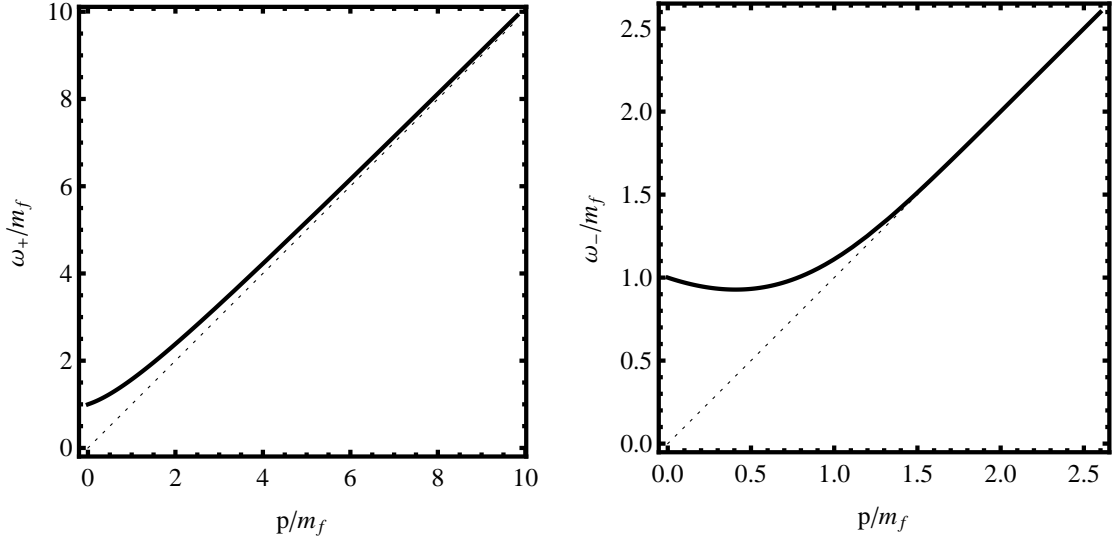}
\caption{Lowest-order quark energies. Ultra-relativistic behavior sets in
quickly.}
\label{quark-energies}
\end{figure}

The quantities $\omega _{\pm }\left( p\right) $ are the solutions to the
lowest-order dispersion relations in which only $\Sigma _{\mathrm{HTL}\pm
}\left( P\right) $ are retained in (\ref{helicity-projected-dispersions}).
These latter are known: 
\begin{equation}
\Sigma _{\mathrm{HTL}\pm }\left( \omega ,p\right) =\frac{m_{f}^{2}}{p}\left[
\pm 1+\frac{1}{2}\left( 1\mp \frac{\omega }{p}\right) \ln \frac{\omega +p}{
\omega -p}\right] ,  \label{Sigma_HTLpm}
\end{equation}
where $m_{f}=\sqrt{C_{F}/8}gT$ is the quark thermal mass to lowest order
with $C_{F}=\left( N_{c}^{2}-1\right) /2N_{c}$. The lowest-order quark
energies $\omega _{\pm }\left( p\right) $ are real; they are displayed in
Fig.~\ref{quark-energies}. Note how quickly the ultra-relativistic behavior
sets in, at already $p\sim 6\left( 2\right) m_{f}$ for $\omega _{+(-)}(p)$.
This indicates that the soft region is effectively narrow. For soft $\bar{p}
\equiv p/m_{f}<1$, they can be obtained in power series: 
\begin{equation}
\omega _{\pm }(p)=m_{f}\left( 1\pm \frac{1}{3}\bar{p}+\frac{1}{3}\bar{p}
^{2}\mp \frac{16}{135}\bar{p}^{3}+\mathcal{O}\left( \bar{p}^{4}\right)
\right) .  \label{lowest-order-quark-energies}
\end{equation}
Also, from the definition of the HTL self-energies $\Sigma _{\mathrm{HTL}}$,
one can rewrite: 
\begin{equation}
\Omega _{\pm }^{(1)}(p)=\frac{\omega _{\pm }^{2}\left( p\right) -p^{2}}{
2m_{f}^{2}}\Sigma _{\pm }^{(1)}\left( \omega _{\pm }(p),p\right) .
\label{Omega-1--rewritten}
\end{equation}

The HTL self-energies $\Sigma _{\mathrm{HTL}}$ define also the HTL-dressed
quark propagator, which can also be decomposed into two components: 
\begin{eqnarray}
\Delta \left( P\right)  &=&\gamma _{+p}~\Delta _{-}\left( P\right) +\gamma
_{-p}\Delta _{+}\left( P\right) ;  \notag \\
\Delta _{\pm }^{-1}\left( P\right)  &=&p_{0}\pm p-\frac{m_{f}^{2}}{p}\left[
\mp 1+\frac{1}{2p}m_{f}^{2}\left( p\pm p_{0}\right) \ln \frac{p_{0}+p}{
p_{0}-p}\right] .  \label{fermion-propagator}
\end{eqnarray}
The HTL-dressed gluon propagator is also a quantity we need. In the Landau
gauge\footnote{
The Landau gauge is part of a class of covariant gauges for which the soft
one-loop order corrections to the lowest-order dispersion relations are
independent of the gauge \cite{schulz}.}, it is given by the following
relation: 
\begin{equation}
\Delta _{\mu \nu }\left( K\right) =\Delta _{T}\left( K\right) P_{\mu \nu
}^{T}+\Delta _{L}\left( K\right) P_{\mu \nu }^{L},  \label{gluon-propagator}
\end{equation}
in which $P_{\mu \nu }^{T,L}$ are the usual transverse and longitudinal
projectors respectively: 
\begin{equation}
P_{\mu \nu }^{T}=g_{\mu \nu }+\frac{\tilde{K}_{\mu }\tilde{K}_{\nu }}{K^{2}}-
\frac{K_{\mu }K_{\nu }}{K^{2}};\quad P_{\mu \nu }^{L}=-\frac{\tilde{K}_{\mu }
\tilde{K}_{\nu }}{K^{2}},  \label{trans-long-projectors}
\end{equation}
where, in the plasma rest-frame, $\tilde{K}=\left( k,k_{0}\hat{k}\right) $.
The quantities $\Delta _{T,L}$ are the transverse and longitudinal gluon
HTL-dressed propagators respectively, given by: 
\begin{eqnarray}
\Delta _{T}^{-1}\left( K\right)  &=&K^{2}-m_{g}^{2}\left[ 1+\frac{K^{2}}{
k^{2}}\left( 1-\frac{k_{0}}{2k}\ln \frac{k_{0}+k}{k_{0}-k}\right) \right] ; 
\notag \\
\Delta _{L}^{-1}\left( K\right)  &=&\left[ K^{2}+2m_{g}^{2}\frac{K^{2}}{k^{2}
}\left( 1-\frac{k_{0}}{2k}\ln \frac{k_{0}+k}{k_{0}-k}\right) \right] .
\label{trans-long-gluon-propagators}
\end{eqnarray}
In this expression, $m_{g}=\sqrt{(N_{c}+N_{f}/2)/6}gT$ is the gluon thermal
mass to lowest order.

\section{The NLO quark self-energy}

\begin{figure}[tbh]
\centering
\includegraphics[width=3.6372in,height=1.2877in]{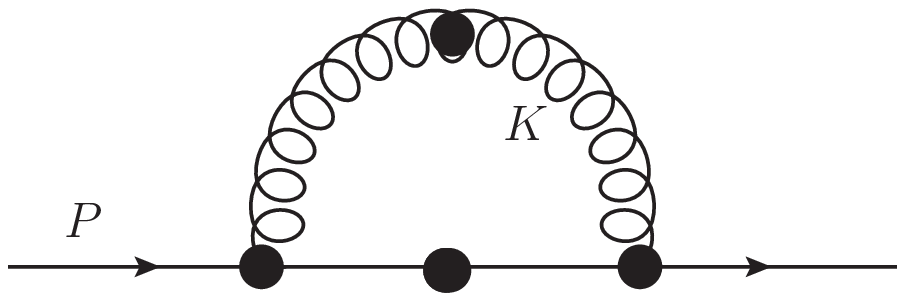}
\caption{NLO HTL-summed quark self-energy $\Sigma _{1}^{(1)}$. The large
dots indicate HTL-dressed vertex functions and propagators. All momenta are
soft.}
\label{NLO-quark-self-energy-1}
\end{figure}

\begin{figure}[tbh]
\centering
\includegraphics[width=2.8609in,height=1.6679in]{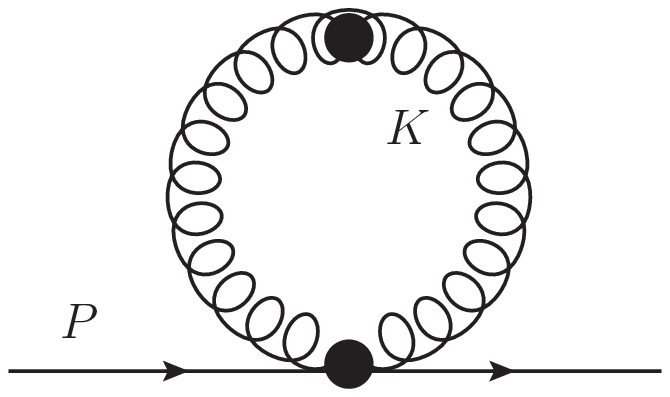}
\caption{NLO HTL-summed quark self-energy $\Sigma _{1}^{(2)}$. The large
dots indicate HTL-dressed vertex functions and propagators. All momenta are
soft.}
\label{NLO-quark-self-energy-2}
\end{figure}

There are two one-loop HTL-dressed diagrams that contribute to the NLO quark
self-energy $\Sigma ^{(1)}$, displayed \cite{axodraw} in Fig.~\ref
{NLO-quark-self-energy-1} and Fig.~\ref{NLO-quark-self-energy-2}. 
The diagram in Fig.~\ref{NLO-quark-self-energy-1} writes as follows: 
\begin{equation}
\Sigma _{1}^{(1)}\left( P\right) =-ig^{2}C_{F}\int \frac{d^{4}K}{\left( 2\pi
\right) ^{4}}\,\Gamma ^{\mu }\left( P,Q\right) \Delta \left( Q\right) \Gamma
^{\nu }\left( Q,P\right) \Delta _{\mu \nu }\left( K\right) ,  \label{Sigma1}
\end{equation}
with $Q=P-K$, and the diagram in Fig.~\ref{NLO-quark-self-energy-2} writes
as: 
\begin{equation}
\Sigma _{2}^{(1)}\left( P\right) =\frac{-ig^{2}C_{F}}{2}\int \frac{d^{4}K}{
\left( 2\pi \right) ^{4}}\mathbf{\,}\Gamma ^{\mu \nu }\left( P,K\right)
\Delta _{\mu \nu }\left( K\right) .  \label{Sigma2}
\end{equation}
There are three summation structures: Lorentz (explicit), Dirac, and RTF. We
introduce now the Keldysh indices (``r/a'' basis) of the closed-time-path
(CTP) formulation of the finite-temperature real-time formalism \cite
{keldysh, real-time-formalism}. The retarted (R), advanced (A), and
symmetric (S) propagators are given by the following definitions: 
\begin{eqnarray}
\Delta _{B,F}^{\mathrm{R}}\left( K\right) &\equiv &\Delta _{B,F}^{\mathrm{ra}
}\left( K\right) =\Delta _{B,F}\left( k_{0}+i\epsilon ,\vec{k}\right) ; 
\notag \\
\Delta _{B,F}^{\mathrm{A}}\left( K\right) &\equiv &\Delta _{B,F}^{\mathrm{ar}
}\left( K\right) =\Delta _{B,F}\left( k_{0}-i\epsilon ,\vec{k}\right) ; 
\notag \\
\Delta _{B,F}^{\mathrm{S}}\left( K\right) &\equiv &\Delta _{B,F}^{\mathrm{rr}
}\left( K\right) =N_{B,F}\left( k_{0}\right) \,\mathrm{sign}(k_{0})\left[
\Delta _{B,F}^{\mathrm{R}}\left( K\right) -\Delta _{B,F}^{\mathrm{A}}\left(
K\right) \right] ,  \label{ret-adv-sy-propa}
\end{eqnarray}
where $B$ stands for bosons and $F$ for fermions, and $N_{B,F}$ are related
to the Bose-Einstein Fermi-Dirac distributions $n_{B,F}$ via the relations: 
\begin{equation}
N_{B,F}\left( x\right) =1\pm 2n_{B,F}\left( x\right) ;\quad n_{B,F}\left(
x\right) =\frac{1}{e^{\lvert x\rvert /T}\mp 1}.
\label{thermal-distributions}
\end{equation}
We then have for the two components of $\Sigma _{1}^{(1)}$ the following
explicit expressions: 
\begin{eqnarray}
\Sigma _{1\pm }^{(1)}\left( P\right) &=&\frac{-ig^{2}C_{F}}{2}\int \frac{
d^{4}K}{\left( 2\pi \right) ^{4}}\mathbf{tr}\mathrm{\,}\gamma _{\pm p}\left[
\Gamma _{\mathrm{arr}}^{\mu }\left( P,Q\right) \Delta ^{\mathrm{R}}\left(
Q\right) \Gamma _{\mathrm{arr}}^{\nu }\left( Q,P\right) \Delta _{\mu \nu }^{
\mathrm{S}}\left( K\right) \right.  \notag \\
&&+\Gamma _{\mathrm{arr}}^{\mu }\left( P,Q\right) \Delta ^{\mathrm{S}}\left(
Q\right) \Gamma _{\mathrm{rar}}^{\nu }\left( Q,P\right) \Delta _{\mu \nu }^{
\mathrm{A}}\left( K\right)  \notag \\
&&+\Gamma _{\mathrm{arr}}^{\mu }\left( P,Q\right) \Delta ^{\mathrm{R}}\left(
Q\right) \Gamma _{\mathrm{aar}}^{\nu }\left( Q,P\right) \Delta _{\mu \nu }^{
\mathrm{A}}\left( K\right)  \notag \\
&&+\Gamma _{\mathrm{aar}}^{\mu }\left( P,Q\right) \Delta ^{\mathrm{R}}\left(
Q\right) \Gamma _{\mathrm{arr}}^{\nu }\left( Q,P\right) \Delta _{\mu \nu }^{
\mathrm{R}}\left( K\right)  \notag \\
&&+\left. \Gamma _{\mathrm{ara}}^{\mu }\left( P,Q\right) \Delta ^{\mathrm{A}
}\left( Q\right) \Gamma _{\mathrm{rar}}^{\nu }\left( Q,P\right) \Delta _{\mu
\nu }^{\mathrm{A}}\left( K\right) \right] ,  \label{Sigma_1pm}
\end{eqnarray}
and for the two components of $\Sigma _{2}^{(1)}$ the following expressions: 
\begin{eqnarray}
\Sigma _{2\pm }^{(1)}\left( P\right) &=&\frac{-ig^{2}C_{F}}{4}\int \frac{
d^{4}K}{\left( 2\pi \right) ^{4}}\mathbf{tr}\,\gamma _{\pm p}\left[ \Gamma _{
\mathrm{arrr}}^{\mu \nu }\left( P,K\right) \Delta _{\mu \nu }^{\mathrm{S}
}\left( K\right) \right.  \notag \\
&&+\left. \Gamma _{\mathrm{aarr}}^{\mu \nu }\left( P,K\right) \Delta _{\mu
\nu }^{\mathrm{R}}\left( K\right) +\Gamma _{\mathrm{arar}}^{\mu \nu }\left(
P,K\right) \Delta _{\mu \nu }^{\mathrm{A}}\left( K\right) \right] .
\label{Sigma_2pm}
\end{eqnarray}
Note that these expressions of $\Sigma _{1\pm }^{(1)}$ in Eq. (\ref
{Sigma_1pm}) and $\Sigma _{2}^{(1)}$ in Eq. (\ref{Sigma_2pm}) have been
written in \cite{carrington--PRD75-2007-045019} using a different notation.

The HTL-dressed vertex functions $\Gamma $ are derived in the literature 
\cite
{defu-heinz--EPJC7-1999-101,defu-et-al--PRD61-2000-085013,fueki-nakkagawa-yokota-yoshida}
. However, there are discrepancies in these results \cite
{defu-et-al--PRD61-2000-085013,fueki-nakkagawa-yokota-yoshida}, which led us
to rederive all three and four-point HTL-dressed vertex functions ab initio
in the CTP formalism. We recover the results\ of \cite
{fueki-nakkagawa-yokota-yoshida}; these are presented in appendix A. For our
needs, we have: 
\begin{eqnarray}
\Gamma _{\mathrm{arr}}^{\mu }\left( P,Q\right) &=&\gamma ^{\mu }+I_{--}^{\mu
}\left( P,Q\right) ;  \notag \\
\Gamma _{\mathrm{rar}}^{\mu }\left( P,Q\right) &=&\gamma ^{\mu }+I_{+-}^{\mu
}\left( P,Q\right) ;  \notag \\
\Gamma _{\mathrm{aar}}^{\mu }\left( P,Q\right) &=&\Gamma _{\mathrm{ara}
}^{\mu }\left( P,Q\right) =0,  \label{3vertices-in-Is}
\end{eqnarray}
for the two-quarks-one-gluon vertices, and: 
\begin{equation}
\Gamma _{\mathrm{arrr}}^{\mu \nu }\left( P,K\right) =I_{--}^{\mu \nu }\left(
P,K\right) ;\quad \Gamma _{\mathrm{aarr}}^{\mu \nu }\left( P,K\right)
=\Gamma _{\mathrm{arar}}^{\mu \nu }\left( P,K\right) =0,
\label{4vertices-in-Is}
\end{equation}
for the two-quarks-two-gluons vertices. In these relations, the quantities $
I $'s are hard thermal loops given by these solid-angle integrals: 
\begin{eqnarray}
I_{\eta _{1}\eta _{2}}^{\mu }\left( P,Q\right) &=&m_{f}^{2}\int \frac{
d\Omega _{s}}{4\pi }\frac{S^{\mu }S\hspace{-7pt}/}{\left( PS+i\eta
_{1}\epsilon \right) \left( QS+i\eta _{2}\epsilon \right) };  \notag \\
I_{\eta _{1}\eta _{2}}^{\mu \nu }\left( P,K\right) &=&m_{f}^{2}\int \frac{
d\Omega _{s}}{4\pi }\frac{-S^{\mu }S^{\nu }S\hspace{-7pt}/}{\left( PS+i\eta
_{1}\epsilon \right) \left( PS+i\eta _{2}\epsilon \right) }  \notag \\
&&\times \left[ \frac{1}{\left( P+K\right) S+i\eta _{1}\epsilon }+\frac{1} {
\left( P-K\right) S+i\eta _{2}\epsilon }\right] ,  \label{the-Is-definition}
\end{eqnarray}
where $S\equiv \left( 1,\hat{s}\right) $ and the indices $\eta _{1}$ and $
\eta _{2}$ take the values $+$ or $-$. Using these results, we can rewrite $
\Sigma _{1\pm }^{(1)}$ and $\Sigma _{2\pm }^{(1)}$ above as simply: 
\begin{eqnarray}
\Sigma _{1\pm }^{(1)}\left( P\right) &=&\frac{-ig^{2}C_{F}}{2}\int \frac{
d^{4}K}{\left( 2\pi \right) ^{4}}\mathbf{tr}\mathrm{\,}\gamma _{\pm p}\left[
\left( \gamma ^{\mu }+I_{--}^{\mu }\left( P,Q\right) \right) \Delta ^{ 
\mathrm{R}}\left( Q\right) \left( \gamma ^{\nu }+I_{--}^{\nu }\left(
Q,P\right) \right) \Delta _{\mu \nu }^{\mathrm{S}}\left( K\right) \right. 
\notag \\
&&+\left( \gamma ^{\mu }+I_{--}^{\mu }\left( P,Q\right) \right) \Delta ^{
\mathrm{S}}\left( Q\right) \left( \gamma ^{\nu }+I_{+-}^{\nu }\left(
Q,P\right) \right) \Delta _{\mu \nu }^{\mathrm{A}}\left( K\right) ;
\label{Sigma1-bis} \\
\Sigma _{2\pm }^{(1)}\left( P\right) &=&\frac{-ig^{2}C_{F}}{4}\int \frac{
d^{4}K}{\left( 2\pi \right) ^{4}}\mathbf{tr}\,\gamma _{\pm p}I_{--}^{\mu \nu
}\left( P,K\right) \Delta _{\mu \nu }^{\mathrm{S}}\left( K\right) .
\label{Sigma2-bis}
\end{eqnarray}

We now work out the contractions over the Dirac and Lorentz indices.
Starting with $\Sigma ^{(1)}$, the contributions that do not include a hard
thermal loop write generically as: 
\begin{eqnarray}
F_{\epsilon _{p};0}^{\mathrm{XY}}\left( P,K\right) &\equiv &\mathbf{tr}
\left( \gamma _{\epsilon _{p}}\gamma ^{\mu }\gamma _{\epsilon _{q}}\gamma
^{\nu }\right) \Delta _{\mu \nu }^{\mathrm{X}}\left( K\right) \Delta
_{-\epsilon _{q}}^{\mathrm{Y}}\left( Q\right)  \notag \\
&=&-2\left( 1-\hat{p}_{\epsilon }.\hat{k}\,\hat{q}_{\epsilon }.\hat{k}
\right) \Delta _{T}^{\mathrm{X}}\left( K\right) \Delta _{-\epsilon _{q}}^{ 
\mathrm{Y}}\left( Q\right)  \notag \\
&&-\left[ k_{0}^{2}\left( 1-\left( \hat{p}_{\epsilon }.\hat{q}_{\epsilon } -2
\hat{p}_{\epsilon }.\hat{k}\,\hat{q}_{\epsilon }.\hat{k}\right) \right)
-2k_{0}k\right.  \notag \\
&&\left. \times \left( \hat{p}_{\epsilon }.\hat{k}+\hat{q}_{\epsilon }.\hat{k
} \right) +k^{2}\left( 1+\hat{p}_{\epsilon }.\hat{q}_{\epsilon }\right) 
\right] \tilde{\Delta}_{L}^{\mathrm{X}}\left( K\right) \Delta _{-\epsilon
_{q}}^{\mathrm{Y}}\left( Q\right) .  \label{trace-no-HTL}
\end{eqnarray}
In this expression, $\tilde{\Delta}_{L}\left( K\right) =\Delta _{L}\left(
K\right) /K^{2}$. Also, $\hat{p}_{\epsilon }=\epsilon _{p}\hat{p}$ with $
\epsilon _{p}=\pm $ and similarly for $\hat{q}_{\epsilon }$, with summation
understood over $\epsilon _{q}$. The superscripts X and Y indicate the RTF
indices (R,\ S, and A). The contributions that involve one hard thermal loop
vertex function writes: 
\begin{eqnarray}
F_{\epsilon _{p};\eta _{1}\eta _{2}}^{\mathrm{XY}}\left( P,K\right) &\equiv
& \mathbf{tr}\left( \gamma _{\epsilon _{p}}I_{\eta _{1}\eta _{2}}^{\mu
}\gamma _{\epsilon _{q}}\gamma ^{\nu }\right) \Delta _{\mu \nu }^{\mathrm{X}
}\left( K\right) \Delta _{-\epsilon _{q}}^{\mathrm{Y}}\left( Q\right)
=m_{f}^{2} \hspace{-4pt}\int \frac{d\Omega _{s}}{4\pi }\frac{1}{\left(
PS+i\eta _{1}\varepsilon \right) \left( QS+i\eta _{2}\varepsilon \right) } 
\notag \\
&&\hspace{-1.2in}\times \left[ \left( 1-\hat{p}_{\epsilon }.\hat{q}
_{\epsilon }-\hat{p}_{\epsilon }.\hat{s}-\hat{q}_{\epsilon }.\hat{s}+\hat{p}
_{\epsilon }.\hat{k}\,\hat{k}.\hat{s}+\hat{q}_{\epsilon }.\hat{k}\,\hat{k}. 
\hat{s}-\hat{k}.\hat{s}^{2}+\hat{p}_{\epsilon }.\hat{q}_{\epsilon }\hat{k}. 
\hat{s}^{2}+2\hat{p}_{\epsilon }.\hat{s}\,\hat{q}_{\epsilon }.\hat{s}\right.
\right.  \notag \\
&&\hspace{-1.2in}\left. -\hat{p}_{\epsilon }.\hat{k}\,\hat{q}_{\epsilon }. 
\hat{s}\,\hat{k}.\hat{s}-\hat{q}_{\epsilon }.\hat{k}\,\hat{p}_{\epsilon }. 
\hat{s}\,\hat{k}.\hat{s}\right) \Delta _{T}^{\mathrm{X}}\left( K\right)
\Delta _{-\epsilon _{q}}^{\mathrm{Y}}\left( Q\right) -\left( k^{2}\left( 1+ 
\hat{p}_{\epsilon }.\hat{q}_{\epsilon }-\hat{p}_{\epsilon }.\hat{s}-\hat{q}
_{\epsilon }.\hat{s}\right) \right.  \label{trace-one-HTL} \\
&&\hspace{-1.2in}-k_{0}k\left( \hat{p}_{\epsilon }.\hat{k}\,+\hat{q}
_{\epsilon }.\hat{k}+2\hat{p}_{\epsilon }.\hat{q}_{\epsilon }\hat{k}.\hat{s}
- \hat{q}_{\epsilon }.\hat{k}\,\hat{p}_{\epsilon }.\hat{s}-\hat{p}_{\epsilon
}. \hat{k}\,\hat{q}_{\epsilon }.\hat{s}-\hat{q}_{\epsilon }.\hat{s}\,\hat{k}
. \hat{s}-\hat{p}_{\epsilon }.\hat{s}\,\hat{k}.\hat{s}\right)  \notag \\
&&\hspace{-1.2in}\left. \left. \hspace{-2pt}+k_{0}^{2}\hspace{-2pt}\left( 
\hat{p}_{\epsilon }.\hat{k}\,\hat{k}.\hat{s}+\hat{q}_{\epsilon }.\hat{k}\, 
\hat{k}.\hat{s}-\hat{k}.\hat{s}^{2}+\hat{p}_{\epsilon }.\hat{q}_{\epsilon } 
\hat{k}.\hat{s}^{2}-\hat{p}_{\epsilon }.\hat{k}\,\hat{q}_{\epsilon }.\hat{s}
\,\hat{k}.\hat{s}-\hat{q}_{\epsilon }.\hat{k}\,\hat{p}_{\epsilon }.\hat{s}\, 
\hat{k}.\hat{s}\right) \right) \tilde{\Delta}_{L}^{\mathrm{X}}\left(
K\right) \Delta _{-\epsilon _{q}}^{\mathrm{Y}}\left( Q\right) \right] . 
\notag
\end{eqnarray}
Here $\eta _{1}$ and $\eta _{2}$ take the values $\pm $. Due to the symmetry
in $\Delta _{\mu \nu }$, The other contribution with one hard-thermal-loop
vertex function is equal to the one above when changing $\eta _{1}$ into $
\eta _{2}$, namely: 
\begin{equation}
\mathbf{tr}\left( \gamma _{\epsilon _{p}}\gamma ^{\mu }\gamma _{\epsilon
_{q}}I_{\eta _{1}\eta _{2}}^{\nu }\right) \Delta _{\mu \nu }^{\mathrm{X}
}\left( K\right) \Delta _{-\epsilon _{q}}^{\mathrm{Y}}\left( Q\right) = 
\mathbf{tr}\left( \gamma _{\epsilon _{p}}I_{\eta _{2}\eta _{1}}^{\mu }\gamma
_{\epsilon _{q}}\gamma ^{\nu }\right) \Delta _{\mu \nu }^{\mathrm{X}}\left(
K\right) \Delta _{-\epsilon _{q}}^{\mathrm{Y}}\left( Q\right) .
\label{trace-mu-nu}
\end{equation}
The contribution involving two hard-thermal-loop vertex functions is longer.
It generically writes: 
\begin{eqnarray}
F_{\epsilon _{p};\eta _{1}\eta _{2};\eta _{1}^{\prime }\eta _{2}^{\prime
}}^{ \mathrm{XY}}\left( P,K\right) &\equiv &\mathbf{tr}\left( \gamma
_{\epsilon _{p}}I_{\eta _{1}\eta _{2}}^{\mu }\gamma _{\epsilon _{q}}I_{\eta
_{1}^{\prime }\eta _{2}^{\prime }}^{\nu }\right) \Delta _{\mu \nu }^{\mathrm{
\ X}}\left( K\right) \Delta _{-\epsilon _{q}}^{\mathrm{Y}}\left( Q\right) 
\notag \\
&=&m_{f}^{4}\int \frac{d\Omega _{s}}{4\pi }\frac{1}{\left( PS+i\eta
_{1}\varepsilon \right) \left( QS+i\eta _{2}\varepsilon \right) }  \notag \\
&&\hspace{-1.5in}\times \int \frac{d\Omega _{s^{\prime }}}{4\pi }\frac{1}{
\left( PS^{\prime }+i\eta _{2}^{\prime }\varepsilon \right) \left(
QS^{\prime }+i\eta _{1}^{\prime }\varepsilon \right) }\left[ \left( -\hat{s}
. \hat{s}^{\prime }-\hat{p}_{\epsilon }.\hat{q}_{\epsilon }\hat{s}.\hat{s}
^{\prime }+\hat{p}_{\epsilon }.\hat{s}\,\hat{s}.\hat{s}^{\prime }+\hat{q}
_{\epsilon }.\hat{s}\,\hat{s}.\hat{s}^{\prime }+\hat{p}_{\epsilon }.\hat{s}
^{\prime }\,\hat{s}.\hat{s}^{\prime }\right. \right.  \notag \\
&&\hspace{-1.5in}+\hat{q}_{\epsilon }.\hat{s}^{\prime }\,\hat{s}.\hat{s}
^{\prime }-\hat{p}_{\epsilon }.\hat{s}\,\hat{q}_{\epsilon }.\hat{s}^{\prime
}\,\hat{s}.\hat{s}^{\prime }-\hat{q}_{\epsilon }.\hat{s}\,\hat{p}_{\epsilon
}.\hat{s}^{\prime }\,\hat{s}.\hat{s}^{\prime }-\hat{s}.\hat{s}^{\prime 2}+ 
\hat{p}_{\epsilon }.\hat{q}_{\epsilon }\hat{s}.\hat{s}^{\prime 2}+\hat{k}. 
\hat{s}\,\hat{k}.\hat{s}^{\prime }+\hat{p}_{\epsilon }.\hat{q}_{\epsilon } 
\hat{k}.\hat{s}\,\hat{k}.\hat{s}^{\prime }  \notag \\
&&\hspace{-1.5in}-\hat{p}_{\epsilon }.\hat{s}\,\hat{k}.\hat{s}\,\hat{k}.\hat{
s}^{\prime }-\hat{q}_{\epsilon }.\hat{s}\,\hat{k}.\hat{s}\,\hat{k}.\hat{s}
^{\prime }-\hat{p}_{\epsilon }.\hat{s}^{\prime }\hat{k}.\hat{s}\,\hat{k}. 
\hat{s}^{\prime }-\hat{q}_{\epsilon }.\hat{s}^{\prime }\hat{k}.\hat{s}\,\hat{
k}.\hat{s}^{\prime }+\hat{p}_{\epsilon }.\hat{s}\,\hat{q}_{\epsilon }.\hat{s}
^{\prime }\hat{k}.\hat{s}\,\hat{k}.\hat{s}^{\prime }  \notag \\
&&\hspace{-1.5in}\left. +\hat{q}_{\epsilon }.\hat{s}\,\hat{p}_{\epsilon }. 
\hat{s}^{\prime }\hat{k}.\hat{s}\,\hat{k}.\hat{s}^{\prime }+\hat{s}.\hat{s}
^{\prime }\hat{k}.\hat{s}\,\hat{k}.\hat{s}^{\prime }-\hat{p}_{\epsilon }. 
\hat{q}_{\epsilon }\hat{s}.\hat{s}^{\prime }\hat{k}.\hat{s}\,\hat{k}.\hat{s}
^{\prime }\right) \Delta _{T}^{\mathrm{X}}\left( K\right) \Delta _{-\epsilon
_{q}}^{\mathrm{Y}}\left( Q\right)  \notag \\
&&\hspace{-1.5in}-\left[ k^{2}\left( 1+\hat{p}_{\epsilon }.\hat{q}_{\epsilon
}-\hat{p}_{\epsilon }.\hat{s}-\hat{q}_{\epsilon }.\hat{s}-\hat{p}_{\epsilon
}.\hat{s}^{\prime }-\hat{q}_{\epsilon }.\hat{s}^{\prime }+\hat{p}_{\epsilon
}.\hat{s}\,\hat{q}_{\epsilon }.\hat{s}^{\prime }+\hat{q}_{\epsilon }.\hat{s}
\,\hat{p}_{\epsilon }.\hat{s}^{\prime }+\hat{s}.\hat{s}^{\prime }-\hat{p}
_{\epsilon }.\hat{q}_{\epsilon }\hat{s}.\hat{s}^{\prime }\right) \right. 
\notag \\
&&\hspace{-1.5in}-k_{0}k\left( \hat{k}.\hat{s}+\hat{k}.\hat{s}^{\prime }+ 
\hat{p}_{\epsilon }.\hat{q}_{\epsilon }\,\hat{k}.\hat{s}+\hat{p}_{\epsilon
}. \hat{q}_{\epsilon }\,\hat{k}.\hat{s}^{\prime }-\hat{p}_{\epsilon }.\hat{s}
\, \hat{k}.\hat{s}-\hat{p}_{\epsilon }.\hat{s}\,\hat{k}.\hat{s}^{\prime }-
\hat{q }_{\epsilon }.\hat{s}\,\hat{k}.\hat{s}-\hat{q}_{\epsilon }.\hat{s}\,
\hat{k}. \hat{s}^{\prime }\right.  \notag \\
&&\hspace{-1.5in}-\hat{p}_{\epsilon }.\hat{s}^{\prime }\,\hat{k}.\hat{s}- 
\hat{p}_{\epsilon }.\hat{s}^{\prime }\,\hat{k}.\hat{s}^{\prime }-\hat{q}
_{\epsilon }.\hat{s}^{\prime }\,\hat{k}.\hat{s}-\hat{q}_{\epsilon }.\hat{s}
^{\prime }\,\hat{k}.\hat{s}^{\prime }+\hat{p}_{\epsilon }.\hat{s}\,\hat{q}
_{\epsilon }.\hat{s}^{\prime }\,\hat{k}.\hat{s}+\hat{p}_{\epsilon }.\hat{s}
\, \hat{q}_{\epsilon }.\hat{s}^{\prime }\,\hat{k}.\hat{s}^{\prime }  \notag
\\
&&\hspace{-1.5in}\left. +\hat{q}_{\epsilon }.\hat{s}\,\hat{p}_{\epsilon }. 
\hat{s}^{\prime }\,\hat{k}.\hat{s}+\hat{q}_{\epsilon }.\hat{s}\,\hat{p}
_{\epsilon }.\hat{s}^{\prime }\,\hat{k}.\hat{s}^{\prime }+\hat{k}.\hat{s}\, 
\hat{s}.\hat{s}^{\prime }+\hat{k}.\hat{s}^{\prime }\,\hat{s}.\hat{s}^{\prime
}-\hat{p}_{\epsilon }.\hat{q}_{\epsilon }\hat{k}.\hat{s}\,\hat{s}.\hat{s}
^{\prime }-\hat{p}_{\epsilon }.\hat{q}_{\epsilon }\hat{k}.\hat{s}^{\prime
}\, \hat{s}.\hat{s}^{\prime }\right)  \label{trace-two-htl} \\
&&\hspace{-1.5in}+k_{0}^{2}\left( \hat{k}.\hat{s}\,\hat{k}.\hat{s}^{\prime
}+ \hat{p}_{\epsilon }.\hat{q}_{\epsilon }\hat{k}.\hat{s}\,\hat{k}.\hat{s}
^{\prime }-\hat{p}_{\epsilon }.\hat{s}\,\hat{k}.\hat{s}\,\hat{k}.\hat{s}
^{\prime }-\hat{q}_{\epsilon }.\hat{s}\,\hat{k}.\hat{s}\,\hat{k}.\hat{s}
^{\prime }-\hat{p}_{\epsilon }.\hat{s}^{\prime }\,\hat{k}.\hat{s}\,\hat{k}. 
\hat{s}^{\prime }-\hat{q}_{\epsilon }.\hat{s}^{\prime }\,\hat{k}.\hat{s}\, 
\hat{k}.\hat{s}^{\prime }\right.  \notag \\
&&\hspace{-1.5in}\left. \left. \left. +\hat{p}_{\epsilon }.\hat{s}\,\hat{q}
_{\epsilon }.\hat{s}^{\prime }\,\hat{k}.\hat{s}\,\hat{k}.\hat{s}^{\prime }+ 
\hat{q}_{\epsilon }.\hat{s}\,\hat{p}_{\epsilon }.\hat{s}^{\prime }\,\hat{k}. 
\hat{s}\,\hat{k}.\hat{s}^{\prime }+\hat{k}.\hat{s}\,\hat{k}.\hat{s}^{\prime
}\,\hat{s}.\hat{s}^{\prime }+\hat{p}_{\epsilon }.\hat{q}_{\epsilon }\hat{k}. 
\hat{s}\,\hat{k}.\hat{s}^{\prime }\,\hat{s}.\hat{s}^{\prime }\right) \right] 
\tilde{\Delta}_{L}^{\mathrm{X}}\left( K\right) \Delta _{-\epsilon _{q}}^{ 
\mathrm{Y}}\left( Q\right) \right] .  \notag
\end{eqnarray}

The integrand in $\Sigma _{2}^{\left( 1\right) }$ is faster to write: 
\begin{eqnarray}
G_{\epsilon _{p};\eta _{1}\eta _{2}}^{\mathrm{X}}\left( P,K\right) &\equiv & 
\frac{1}{2}\mathbf{tr}\left( \gamma _{\epsilon _{p}}I_{\eta _{1}\eta
_{2}}^{\mu \nu }\right) \Delta _{\mu \nu }^{\mathrm{X}}\left( K\right)
=m_{f}^{2}\int \frac{d\Omega _{s}}{4\pi }\frac{1}{\left[ PS+i\eta
_{1}\varepsilon \right] \left[ PS+i\eta _{2}\varepsilon \right] }  \notag \\
&&\times \left[ \frac{1}{\left( P+K\right) S+i\eta _{1}\varepsilon }+\frac{1
} {\left( P-K\right) S+i\eta _{2}\varepsilon }\right]  \notag \\
&&\times \left( 1-\hat{p}_{\epsilon }.\hat{s}-\hat{k}.\hat{s}^{2}+\hat{p}
_{\epsilon }.\hat{s}\,\hat{k}.\hat{s}^{2}\right) \Delta _{T}^{\mathrm{X}
}\left( K\right)  \label{trace-in-Sigma2} \\
&&\left( k^{2}\left( 1-\hat{p}_{\epsilon }.\hat{s}\right) -2k_{0}k\left( 
\hat{k}.\hat{s}-\hat{p}_{\epsilon }.\hat{s}\,\hat{k}.\hat{s}\right)
+k_{0}^{2}\,\hat{p}_{\epsilon }.\hat{s}\,\hat{k}.\hat{s}^{2}\right) \tilde{
\Delta}_{L}^{\mathrm{X}}\left( K\right) .  \notag
\end{eqnarray}
From these expressions, we can write the NLO HTL-dresses quark self-energy
in a compact form: 
\begin{eqnarray}
\Sigma _{\pm }^{(1)}\left( P\right) &=&\frac{-ig^{2}C_{F}}{2}\int \frac{
d^{4}K}{\left( 2\pi \right) ^{4}}\left[ F_{\pm ;0}^{\mathrm{SR}}\left(
P,K\right) +F_{\pm ;0}^{\mathrm{AS}}\left( P,K\right) +2F_{\pm ;--}^{\mathrm{
\ SR}}\left( P,K\right) +F_{\pm ;--}^{\mathrm{AS}}\left( P,K\right) \right .
\notag \\
&&\left. +F_{\pm ;-+}^{\mathrm{AS}}\left( P,K\right) +F_{\pm ;--;--}^{ 
\mathrm{SR}}\left( P,K\right) +F_{\pm ;--;+-}^{\mathrm{AS}}\left( P,K\right)
+G_{\pm ;--}^{\mathrm{S}}\left( P,K\right) \right] .  \label{Sigma-compact}
\end{eqnarray}

\section{HTL vertex functions and Feynman parametrization}

The next step is to find a way to evaluate the solid-angle integrals
involved in the hard-thermal-loop vertex functions. The way we do this is to
rewrite these integrals using the Feynman parametrization. From Eqs.~(\ref
{trace-one-HTL}), (\ref{trace-two-htl}), and (\ref{trace-in-Sigma2}) above,
we see that we have two kinds of solid-angle integrals to deal with, namely: 
\begin{eqnarray}
J_{\eta _{1}\eta _{2}}^{\mu \alpha }\left( P,Q\right) &=&\int \frac{d\Omega
_{s}}{4\pi }\frac{S^{\mu }S^{\alpha }}{\left[ PS+i\eta _{1}\varepsilon 
\right] \left[ QS+i\eta _{2}\varepsilon \right] };  \notag \\
I_{\eta _{1}\eta _{2}}^{\mu \nu \alpha }\left( P,K\right) &=&\int \frac{
d\Omega _{s}}{4\pi }\frac{S^{\mu }S^{\nu }S^{\alpha }}{\left[ PS+i\eta
_{1}\varepsilon \right] \left[ PS+i\eta _{2}\varepsilon \right] }  \notag \\
&&\times \left[ \frac{1}{\left( P+K\right) S+i\eta _{1}\varepsilon }+\frac{1
} {\left( P-K\right) S+i\eta _{2}\varepsilon }\right] .
\label{solid-angle-integrals}
\end{eqnarray}

The simplest of all these integrals is the integral: 
\begin{equation}
J_{\eta _{1}\eta _{2}}^{00}\left( P,Q\right) =\int \frac{d\Omega _{s}}{4\pi }
\frac{1}{\left( PS+i\eta _{1}\varepsilon \right) \left( QS+i\eta
_{2}\varepsilon \right) }.  \label{I-00}
\end{equation}
Remember that $S$ is the 4-vector ($1,\hat{s})$ and the integration is over
the solid angle of $\hat{s}$. Let us put aside the $i\varepsilon $
prescription for the moment. Using the Feynman parametrization: 
\begin{equation}
\frac{1}{AB}=\int_{0}^{1}du\frac{1}{\left( Au+B\left( 1-u\right) \right)
^{2} },  \label{feyman-para1}
\end{equation}
we write: \ 
\begin{equation}
J_{\eta _{1}\eta _{2}}^{00}\left( P,Q\right) =\int_{0}^{1}du\int \frac{
d\Omega _{s}}{4\pi }\frac{1}{\left[ \left( P-Ku\right) S\right] ^{2}}
=\int_{0}^{1}\frac{du}{\left( P-Ku\right) ^{2}},  \label{I-00-FeyPara}
\end{equation}
where $K=P-Q$. In this case, the integral over $u$ can be done formally to
get: 
\begin{equation}
J_{\eta _{1}\eta _{2}}^{00}\left( P,Q\right) =\frac{1}{2\sqrt{\Delta }}\ln 
\frac{\left( 1-u_{1}\right) u_{2}}{\left( 1-u_{2}\right) u_{1}},
\label{I-00-integrated}
\end{equation}
with the notation $u_{1,2}=\left( PK\pm \sqrt{\Delta }\right) /K^{2}$ and $
\Delta =PK^{2}-P^{2}K^{2}$. Noticing that $PS+i\eta _{1}\varepsilon
=p_{0}+i\eta _{1}\varepsilon -\vec{p}.\hat{s}$, the reintroduction of the $
i\varepsilon $'s in the final result (\ref{I-00-integrated}) is a matter of
shifting $p_{0}\rightarrow p_{0}+i\eta _{1}\varepsilon $ and $
q_{0}\rightarrow q_{0}+i\eta _{2}\varepsilon $. This also applies to the two
next integrals.

The next HTL vertex function to consider is the solid-angle integral: 
\begin{equation}
J_{\eta _{1}\eta _{2}}^{0i}\left( P,Q\right) =\int \frac{d\Omega _{s}}{4\pi }
\frac{\hat{s}^{i}}{\left( PS+i\eta _{1}\varepsilon \right) \left( QS+i\eta
_{2}\varepsilon \right) }.  \label{I-0i}
\end{equation}
Using a Feynman parametrization and the notation $R=P-\left( P-Q\right) u$
in which $p_{0}$ and $q_{0}$ are redefined with the corresponding $
i\varepsilon $'s, we obtain the result: 
\begin{eqnarray}
J_{\eta _{1}\eta _{2}}^{0i}\left( P,Q\right) &=&\frac{1}{2}\int_{0}^{1}du 
\frac{r^{i}}{r}\int_{-1}^{1}dx\frac{x}{\left( r_{0}-xr\right) ^{2}}  \notag
\\
&=&\int_{0}^{1}du\left[ \frac{r_{0}}{r_{0}^{2}-r^{2}}-\frac{1}{2r}\ln \frac{
r_{0}+r}{r_{0}-r}\right] \frac{r^{i}}{r^{2}}.  \label{I-0i-integrated}
\end{eqnarray}
Little useful comes from pushing further the integration over $u$ as the
final result will not have a reasonably simple formal expression; it is
better to leave this expression as it is and let the integration over $u$ be
done numerically. However, the Feynman parametrization is useful as it
reduces the number of integrations to be performed from two (solid angle) to
one (over $u$) or zero when this latter is explicitly feasible.

The third HTL-vertex solid-angle integral is: 
\begin{equation}
J_{\eta _{1}\eta _{2}}^{ij}\left( P,Q\right) =\int \frac{d\Omega _{s}}{4\pi }
\frac{\hat{s}^{i}\hat{s}^{j}}{\left( PS+i\eta _{1}\varepsilon \right) \left(
QS+i\eta _{2}\varepsilon \right) }.  \label{I-ij}
\end{equation}
This solid-angle integral can be performed using the Feynman parametrization
and, still with $R=P-\left( P-Q\right) u$, we have: 
\begin{eqnarray}
J_{\eta _{1}\eta _{2}}^{ij}\left( P,Q\right) &=&\int_{0}^{1}du\left( A_{\eta
_{1}\eta _{2}}\delta ^{ij}+B_{\eta _{1}\eta _{2}}\hat{r}^{i}\hat{r}
^{j}\right) ;  \notag \\
A_{\eta _{1}\eta _{2}} &=&-\frac{1}{r^{2}}\left( 1-\frac{r_{0}}{2r}\ln \frac{
r_{0}+r}{r_{0}-r}\right) ;  \notag \\
B_{\eta _{1}\eta _{2}} &=&\frac{1}{r_{0}^{2}-r^{2}}+\frac{3}{r^{2}}\left( 1- 
\frac{r_{0}}{2r}\ln \frac{r_{0}+r}{r_{0}-r}\right) .  \label{I-ij-integrated}
\end{eqnarray}
The integration over $u$ is left to be performed numerically.

The two-gluon-two-fermion HTL vertex function uses the Feynman integral: 
\begin{equation}
\frac{1}{ABC}=2!\int_{0}^{1}du_{1}\int_{0}^{1}du_{2}\frac{u_{1}}{\left[
\left( C-B\right) u_{1}u_{2}+\left( B-A\right) u_{1}+A\right] ^{3}}.
\label{feyman-para2}
\end{equation}
Now calling $P_{1,2}\equiv \left( p_{0}+i\eta _{1,2}\varepsilon ,\vec{p}
\right) $, we rewrite: 
\begin{eqnarray}
I_{\eta _{1}\eta _{2}}^{\mu \nu \alpha }\left( P,K\right) &=&J_{\eta
_{1}\eta _{2}}^{\mu \nu \alpha }\left( P,K\right) +J_{\eta _{2}\eta
_{1}}^{\mu \nu \alpha }\left( P,-K\right) ;  \notag \\
J_{\eta _{1}\eta _{2}}^{\mu \nu \alpha }\left( P,K\right) &=&\int \frac{
d\Omega _{s}}{4\pi }\frac{S^{\mu }S^{\nu }S^{\alpha }}{P_{1}S\,P_{2}S\,
\left( P_{1}+K\right) S}.  \label{4-vertex-solid-angle}
\end{eqnarray}
Calling $C=\left( P_{1}+K\right) S$, $B=P_{1}S$, $A=P_{2}S$, and the
four-vector $T\equiv u_{1}u_{2}K+u_{1}\left( P_{1}-P_{2}\right) +P_{2}$,
with $t_{0}=u_{1}u_{2}k_{0}+iu_{1}\left( \eta _{1}-\eta _{2}\right)
\varepsilon +p_{0}+i\eta _{2}\varepsilon $ and $\vec{t}=u_{1}u_{2}\vec{k}+ 
\vec{p}$, we have: 
\begin{equation}
J_{\eta _{1}\eta _{2}}^{\mu \nu \alpha }\left( P,K\right)
=2\int_{0}^{1}du_{1}\,u_{1}\int_{0}^{1}du_{2}\int \frac{d\Omega _{s}}{4\pi } 
\frac{S^{\mu }S^{\nu }S^{\alpha }}{\left( t_{0}-\vec{t}.\hat{s}\right) ^{3}}.
\label{4vertex-I1}
\end{equation}
Since the quantity $J_{\eta _{1}\eta _{2}}^{\mu \nu \alpha }$ is fully
symmetric in the Lorentz indices, we only need to find the $000$, $00i$, $
0ij $, and $ijk$ components. We have: 
\begin{eqnarray}
J_{\eta _{1}\eta _{2}}^{000}\left( P,K\right)
&=&2\int_{0}^{1}du_{1}\,u_{1}\int_{0}^{1}du_{2}\int \frac{d\Omega _{s}}{4\pi 
}\frac{1}{\left( t_{0}-\vec{t}.\hat{s}\right) ^{3}}  \notag \\
&=&2\int_{0}^{1}du_{1}\,u_{1}\int_{0}^{1}du_{2}\frac{t_{0}}{\left(
t_{0}^{2}-t^{2}\right) ^{2}}.  \label{I-000}
\end{eqnarray}
Note that in this case, the Feynman parametrization does not reduce the
number of integrations to perform. Yet, it still has an advantage over the
original integration over the solid angle $\Omega _{s}$ of the original unit
vector $\hat{s}$ as, cast this way, the subsequent integration over the
azimuthal angle of $\hat{k}$ will now be trivial. The next solid-angle
integral is: 
\begin{eqnarray}
J_{\eta _{1}\eta _{2}}^{00i}\left( P,K\right)
&=&2\int_{0}^{1}du_{1}\,u_{1}\int_{0}^{1}du_{2}\int \frac{d\Omega _{s}}{4\pi 
}\frac{\hat{s}^{i}}{\left( t_{0}-\vec{t}.\hat{s}\right) ^{3}}  \notag \\
&=&2\int_{0}^{1}du_{1}\,u_{1}\int_{0}^{1}du_{2}\frac{t^{i}}{\left(
t_{0}^{2}-t^{2}\right) ^{2}}.  \label{I-00i}
\end{eqnarray}
The next solid-angle integral involves a symmetric rank-2 tensor structure: 
\begin{eqnarray}
J_{\eta _{1}\eta _{2}}^{0ij}\left( P,K\right)
&=&2\int_{0}^{1}du_{1}\,u_{1}\int_{0}^{1}du_{2}\int \frac{d\Omega _{s}}{4\pi 
}\frac{\hat{s}^{i}\hat{s}^{j}}{\left( t_{0}-\vec{t}.\hat{s}\right) ^{3}} 
\notag \\
&=&2\int_{0}^{1}du_{1}\,u_{1}\int_{0}^{1}du_{2}\left( C_{\eta _{1}\eta
_{2}}\delta ^{ij}+D_{\eta _{1}\eta _{2}}\hat{t}^{i}\hat{t}^{j}\right) ; 
\notag \\
C_{\eta _{1}\eta _{2}} &=&\frac{t_{0}}{2t^{2}\left( t_{0}^{2}-t^{2}\right) }
- \frac{1}{4t^{3}}\ln \frac{t_{0}+t}{t_{0}-t};  \notag \\
D_{\eta _{1}\eta _{2}} &=&\frac{t_{0}\left( 5t^{2}-3t_{0}^{2}\right) } {
2t^{2}\left( t_{0}^{2}-t^{2}\right) ^{2}}+\frac{3}{4t^{3}}\ln \frac{t_{0}+t
} {t_{0}-t},  \label{I-0ij}
\end{eqnarray}
and the last solid-angle integral involves a completely symmetric rank-3
tensor structure: 
\begin{eqnarray}
J_{\eta _{1}\eta _{2}}^{ijk}\left( P,K\right)
&=&2\int_{0}^{1}du_{1}\,u_{1}\int_{0}^{1}du_{2}\int \frac{d\Omega _{s}}{4\pi 
}\frac{\hat{s}^{i}\hat{s}^{j}\hat{s}^{k}}{\left( t_{0}-\vec{t}.\hat{s}
\right) ^{3}}  \notag \\
&=&2\int_{0}^{1}du_{1}\,u_{1}\int_{0}^{1}du_{2}\left[ E_{\eta _{1}\eta
_{2}}\left( \hat{t}^{i}\delta ^{jk}+\hat{t}^{j}\delta ^{ki}+\hat{t}
^{k}\delta ^{ij}\right) +F_{\eta _{1}\eta _{2}}\hat{t}^{i}\hat{t}^{j}\hat{t}
^{k}\right] ;  \notag \\
E_{\eta _{1}\eta _{2}} &=&\frac{1}{2t^{3}}\left( 2+\frac{t_{0}^{2}}{\left(
t_{0}^{2}-t^{2}\right) }-\frac{3t_{0}}{2t}\ln \frac{t_{0}+t}{t_{0}-t}\right)
;  \notag \\
F_{\eta _{1}\eta _{2}} &=&\frac{t}{\left( t_{0}^{2}-t^{2}\right) ^{2}}-\frac{
5}{2t^{3}}\left( 2+\frac{t_{0}^{2}}{\left( t_{0}^{2}-t^{2}\right) }-\frac{
3t_{0}}{2t}\ln \frac{t_{0}+t}{t_{0}-t}\right) .  \label{I-ijk}
\end{eqnarray}

\section{Integration: A prototype}

The next task is of course to evaluate all the integrals in Eq.~(\ref
{Sigma-compact}), one after the other, and add them together. This is not an
easy matter. Here we show how one can carry out with such expressions. We
work in units of the quark thermal mass, i.e., we take $m_{f}=1$. This means
that the ratio: 
\begin{equation}
\frac{m_{g}^{2}}{m_{f}^{2}}=\frac{16N_{c}\left( N_{c}+N_{f}/2\right) }{
6\left( N_{c}^{2}-1\right) }=\left( 3+\frac{N_{f}}{2}\right) ,  \label{mg}
\end{equation}
with $N_{c}=3$ will replace $m_{g}^{2}$ in the gluon propagators. This ratio
is equal to 4 for $N_{f}=2$, a value we take in this section.

\begin{figure}[htb]
\centering
\includegraphics[width=4.3495in,height=2.709in]{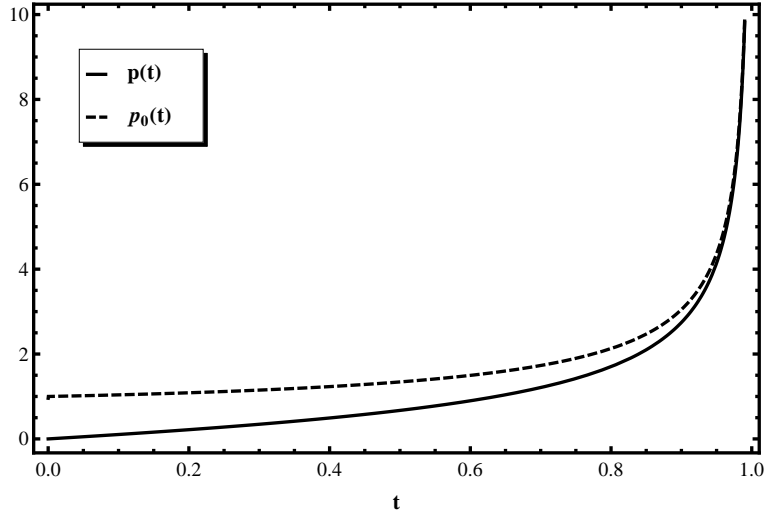}
\caption{The quark momentum $p$ (solid)\ and energy $p_{0}$ (dashed) as
functions of the ratio $t=p/p_{0}$. The quark mass $m_{f}$ is taken equal to
one. The fast-moving limit is $t\rightarrow 1$. Slow-moving quarks would be
for $t\lesssim 0.64$.}
\label{p-wm_t}
\end{figure}

It is best to work out a prototype, namely\footnote{
Prefactors are not important in this section. They are not displayed. The
quantities $\func{Re}I(t,\varepsilon )$ and $\func{Im}I(t,\varepsilon $) are
plotted in arbitrary units.}: 
\begin{eqnarray}
I\left( t,\varepsilon \right)  &=&\int d^{4}K\Delta _{T}^{\mathrm{S}}\left(
K\right) \Delta _{-}^{\mathrm{R}}\left( Q\right) \int \frac{d\Omega _{s}}{
4\pi }\frac{1}{\left( PS-i\varepsilon \right) \left( QS-i\varepsilon \right) 
}  \notag \\
&=&\int_{0}^{+\infty }dk\int_{-1}^{+1}dx\int_{-\infty }^{+\infty
}dk_{0}\int_{0}^{1}du\,M\left( t,k,k_{0},x,u,\varepsilon \right) .
\label{I_t_epsilon}
\end{eqnarray}
The notation is explained as we proceed. As before, $P=(p_{0},\vec{p})$ is
the four-momentum of the external soft "on-shell" quark, $K=(k_{0},\vec{k})$
the four-momentum of the soft gluon in the loop, and $Q=P-K$; see Fig. \ref
{NLO-quark-self-energy-1}. The quantity $x$ is cosine the angle between $
\vec{k}$ and $\vec{p}$ -- the integration over the azimuthal angle of $\hat{k
}$ is done trivially. The integration variable $u$ is from the Feynman
parametrization when the solid-angle integral is performed, see Eq.~(\ref
{I-00-FeyPara}). The quantity $t$ is the ratio $p/p_{0}$, to be introduced
more naturally later in this discussion. The integrand $M$ in this example
is therefore given by the following relation: 
\begin{equation}
M\left( t,k,k_{0},x,u,\varepsilon \right) =\frac{k^{2}N_{B}\left(
k_{0}\right) \mathrm{sign}\left( k_{0}\right) }{Fu^{2}+2Bu+A}\left[ \Delta
_{T}^{\mathrm{R}}\left( k,k_{0},\varepsilon \right) -\Delta _{T}^{\mathrm{R}
}\left( k,k_{0},-\varepsilon \right) \right] \Delta _{-}^{\mathrm{R}}\left(
q,q_{0},\varepsilon \right) .  \label{integrand-M}
\end{equation}
The quantity $N_{B}\left( k_{0}\right) $ is related to the Bose-Einstein
distribution, see Eq.~(\ref{thermal-distributions}), and $\Delta _{T}^{
\mathrm{R}}\left( k,k_{0},\varepsilon \right) $ is the retarded transverse
gluon propagator, which can be written as: 
\begin{eqnarray}
\Delta _{T}^{\mathrm{R}(-1)}\left( k,k_{0},\varepsilon \right)  &=&\frac{
4k_{0}^{2}}{3k^{2}}+\left( k^{2}-k_{0}^{2}\right) \left[ 1-\frac{k_{0}}{
3k^{3}}\ln \left( \frac{(k_{0}-k)^{2}+\varepsilon ^{2}}{(k_{0}+k)^{2}+
\varepsilon ^{2}}\right) \right]   \label{inverse-gluon-propagator} \\
&&\hspace{-0.5in}+\boldsymbol{i}\left[ \frac{2k_{0}}{3k^{2}}\left(
k_{0}^{2}-k^{2}\right) \left( \arctan \left( \frac{\varepsilon }{k_{0}-k}
\right) -\arctan \left( \frac{\varepsilon }{k_{0}+k}\right) \right)
-\varepsilon \,\mathrm{sign}\left( k_{0}\right) \right] .  \notag
\end{eqnarray}
The quantity $\Delta _{-}^{\mathrm{R}}\left( q,q_{0},\varepsilon \right) $
is the retarted quark propagator, given by: 
\begin{eqnarray}
\Delta _{-}^{\mathrm{R}\left( -1\right) }\left( q,q_{0},\varepsilon \right) 
&=&\frac{1}{q}+q-q_{0}-\frac{q_{0}-q}{4q^{2}}\ln \left( \frac{
(q_{0}+q)^{2}+\varepsilon ^{2}}{(q_{0}-q)^{2}+\varepsilon ^{2}}\right)  
\notag \\
&&+\frac{\varepsilon }{2q^{2}}\left[ \arctan \left( \frac{\varepsilon }{
q_{0}+q}\right) -\arctan \left( \frac{\varepsilon }{q_{0}-q}\right) \right] 
\notag \\
&&-\boldsymbol{i}\left( \varepsilon +\frac{\varepsilon }{4q^{2}}\ln \left( 
\frac{(q_{0}+q)^{2}+\varepsilon ^{2}}{(q_{0}-q)^{2}+\varepsilon ^{2}}\right)
\right.   \notag \\
&&+\left. \frac{q_{0}-q}{2q^{2}}\left[ \arctan \left( \frac{\varepsilon }{
q_{0}+q}\right) -\arctan \left( \frac{\varepsilon }{q_{0}-q}\right) \right]
\right) .  \label{inverse-quark-propagator}
\end{eqnarray}
The momentum $\vec{q}=\vec{p}-\vec{k}$ and the energy $q_{0}=p_{0}-k_{0}$
where $p_{0}$ is the quark energy. The quantities $F$, $B$ and $A$ in the
denominator of $M$ are regularized relativistic relations: 
\begin{eqnarray}
A &=&\left( p_{0}-i\varepsilon \right) ^{2}-p^{2};  \notag \\
B &=&\vec{k}.\vec{p}-\left( k_{0}-2i\varepsilon \right) \left(
p_{0}-i\varepsilon \right) ;  \notag \\
F &=&\left( k_{0}-2i\varepsilon \right) ^{2}-k^{2}.  \label{A-B-F}
\end{eqnarray}
As mentioned above, the variable $t$ is the ratio $p/p_{0}$, in terms of
which the quark momentum writes: 
\begin{equation}
p\left( t\right) /m_{f}=\sqrt{\frac{t}{1-t}-\frac{1}{2}\ln \left( \frac{1+t}{
1-t}\right) }.  \label{p_t}
\end{equation}
Remember we take the quark thermal mass $m_{f}=1$. The above relation comes
from the lowest-order quark dispersion relation $\Delta _{-}^{-1}(p_{0},p)=0$
where $\Delta _{-}^{-1}$ is given in Eq. (\ref{fermion-propagator}). The
behavior of $p(t)$ and $p_{0}(t)=\omega _{-}(p(t))$ is shown in Fig.~\ref
{p-wm_t}. Slow-moving quarks would have $p\lesssim 1$ (in units of $m_{f}$),
which translates into $t\lesssim 0.64$. The limit $t\rightarrow 1$ is the
fast-moving region in which both $p(t)$ and $p_{0}(t)$ are large.

In this example, the integral over $u$ can be done analytically,
which reduces the number of integrations in
Eq.~(\ref{I_t_epsilon}) to three. Indeed, one obtains:
\begin{eqnarray}
I(t,\varepsilon ) &=&\int_{0}^{+\infty }dk\int_{-1}^{+1}dx\int_{-\infty
}^{+\infty }dk_{0}\,k^{2}N_{B}\left( k_{0}\right) \mathrm{sign}\left(
k_{0}\right)   \notag \\
&&\times \left[ \Delta _{T}^{\mathrm{R}}\left( k,k_{0},\varepsilon \right)
-\Delta _{T}^{\mathrm{R}}\left( k,k_{0},-\varepsilon \right) \right] \Delta
_{-}^{\mathrm{R}}\left( q,q_{0},\varepsilon \right) V\left(
t,k,k_{0},x,\varepsilon \right) ,  \label{I-t-in-V}
\end{eqnarray}
with the function $V$ given by the following expression,
see Eq.~(\ref{I-00-integrated}):
\begin{eqnarray}
V\left( t,k,k_{0},x,\varepsilon \right)  &=&\int_{0}^{1}du\frac{1}{
Fu^{2}+2Bu+A}  \notag \\
&=&\frac{1}{2\sqrt{\Delta }}\left( \frac{1}{2}\ln \left[ \frac{\left( \left(
r_{1}-1\right) ^{2}+i_{1}{}^{2}\right) \left( r_{2}{}^{2}+i_{2}{}^{2}\right) 
}{\left( \left( r_{2}-1\right) ^{2}+i_{2}{}^{2}\right) \left(
r_{1}^{2}+i_{1}{}^{2}\right) }\right] \right.   \notag \\
&&+\boldsymbol{i}\left[ \arctan \left( \frac{r_{1}}{i_{1}}\right) -\arctan
\left( \frac{r_{1}-1}{i_{1}}\right) \right.   \notag \\
&&\left. -\left. \arctan \left( \frac{r_{2}}{i_{2}}\right) +\arctan \left( 
\frac{r_{2}-1}{i_{2}}\right) \right] \right) .  \label{kernal-V}
\end{eqnarray}
The quantities are $r_{i}=\func{Re}s_{i}$ and $i_{i}=\func{Im}s_{i}$ with: 
\begin{equation}
s_{1}=\frac{-B+\sqrt{\Delta }}{F};\quad s_{2}=\frac{-B-\sqrt{\Delta }}{F}
;\quad \Delta =B^{2}-AF.  \label{s_1-s_2}
\end{equation}

\begin{figure}[tbh]
\centering
\includegraphics[width=4.5886in,height=2.8717in]{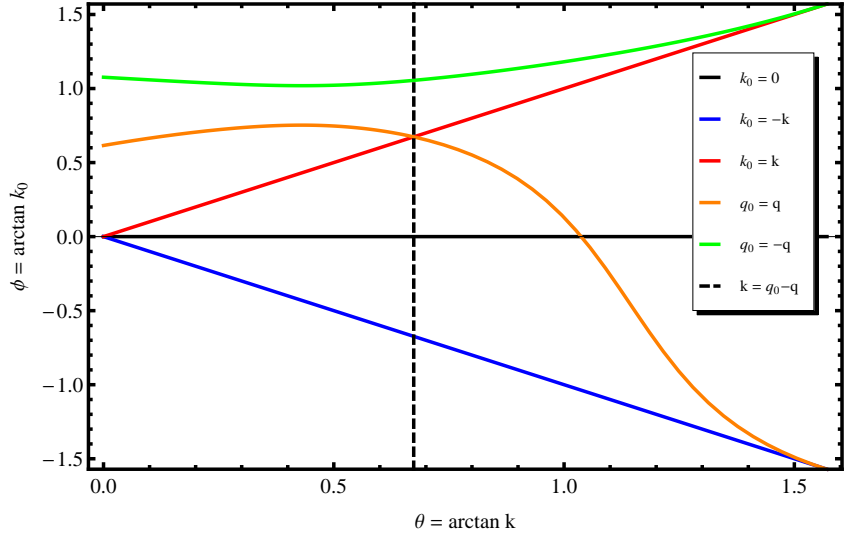}
\caption{Boundaries in $\left( k,k_{0}\right) $ plane at which the integrand 
$M$ has sharp jumps (see text for details). Here $t=0.45$ and $x=0.8$.}
\label{jump-regions-theta-phi-v2}
\end{figure}

\begin{figure}[htb]
\centering
\includegraphics[width=6.699in,height=2.137in]{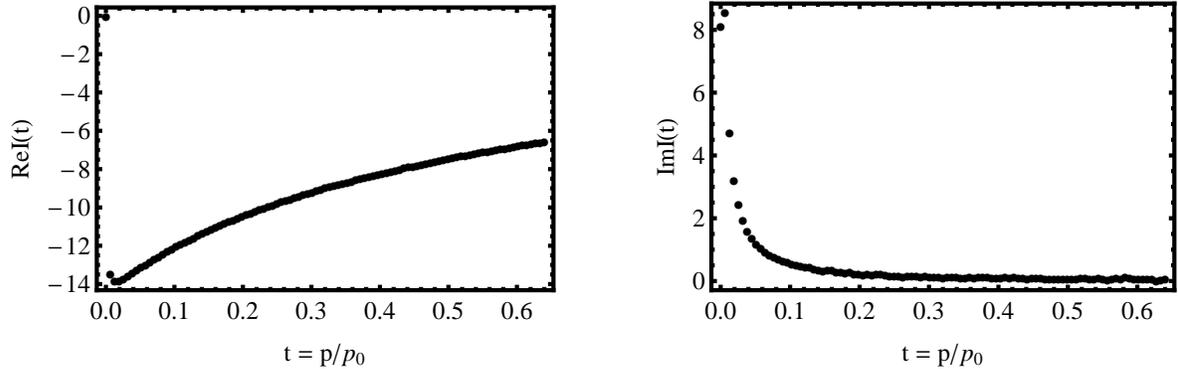}
\caption{The real and imaginary parts of $I(t)$ for soft $p$ (meaning $
t\lesssim 0.64$). Both parts have a stable behavior. Here, $\protect
\varepsilon =3.2\times 10^{-6}$. The units of $I(t)$ are arbitrary.}
\label{I-vs-t-v2}
\end{figure}

There are difficulties with the integral $I\left( t,\varepsilon \right) $.
One is that the integrand features sudden jumps, those coming from the
arctans in the propagators, the Bose-Einstein distribution at $k_{0}=0$, and
eventually those coming from $V$. Such jumps make any integration method
either too long to be useful, or give unstable results. This instability is
more intense when trying to take $\varepsilon $ smaller and smaller. So, in
order to see more closely what is at stake, we have partitioned the
integration region in the $\left( k,k_{0}\right) $ plane into domains
bounded by the following lines of sudden jumps:
\begin{eqnarray}
k_{0} &=&0;\quad k_{0}=\pm k;\quad k_{0}=p_{0}\pm \sqrt{p^{2}+k^{2}-2pkx}; 
\notag \\
k &=&k_{t}\equiv \frac{1}{2}\frac{p_{0}^{2}-p^{2}}{p_{0}-xp}=\frac{1}{2t}
\frac{1-t^{2}}{1-xt}\sqrt{\frac{t}{1-t}-\frac{1}{2}\ln \left( \frac{1+t}{1-t}
\right) }.  \label{jump-lines}
\end{eqnarray}
The last (vertical) line is simply the line $k=p_{0}-q$. Using the change of
variables $\theta =\arctan k$ and $\phi =\arctan k_{0}$, these domains are
plotted in Fig.~\ref{jump-regions-theta-phi-v2}. 
If one integrates $M$ in each of the domains of Fig.~\ref
{jump-regions-theta-phi-v2} separately and sum up the results, one finds
that $I\left( t,\varepsilon \right) $ becomes stable and reliable in the
limit $\varepsilon \rightarrow 0$. See Fig.~\ref{I-vs-t-v2} (the units of $
I(t,\varepsilon )$ there are arbitrary): both the real and imaginary parts
of $I(t)$ behave smoothly, and these two plots are obtained with $
\varepsilon =3.2\times 10^{-6}$ (in units of $m_{f}$).

\begin{figure}[tbh]
\centering
\includegraphics[width=6.699in,height=2.137in]{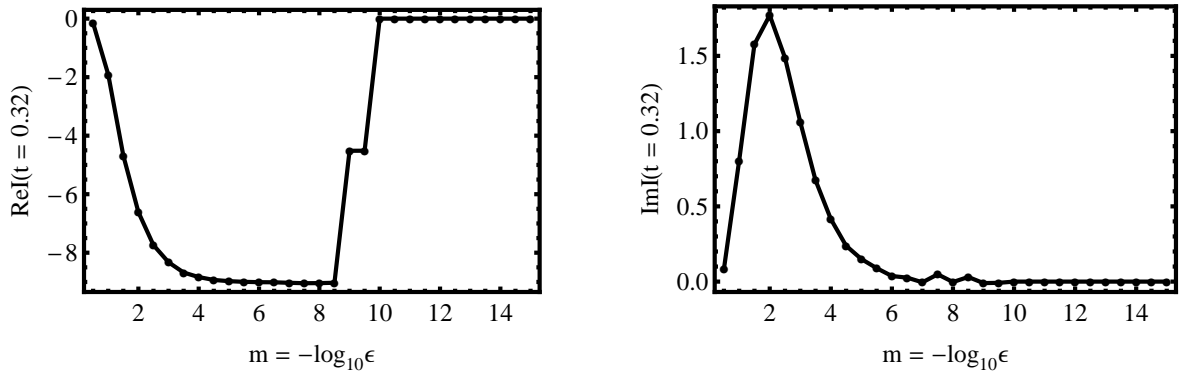}
\caption{The $\protect\varepsilon $-dependence of $I(t)$. We see stability
and convergence up until $\protect\epsilon \simeq 10^{-8}$. Here $t=p/p_{0}$
is take equal to $0.32$.}
\label{I-vs-e}
\end{figure}

The dependence in $\varepsilon $ is of course an issue to explore. Fig.~\ref
{I-vs-e} shows how for example $I(t=0.32,\varepsilon )$ behaves as a
function of $m=-\log _{10}\varepsilon $. 
The behavior is stable down until $\varepsilon \simeq 10^{-8}$ (in units of $
m_{f}$), below which the numerics loose reliability. This is way beyond any
precision we can hope for, and we notice that $I(t,\varepsilon )$ converges
smoothly to a finite value already satisfactorily reached at $\varepsilon
\simeq 10^{-6}$. The smooth convergence to finite values is also obtained
for other values of $t$.

\section{Outlook}

The present work aims at calculating the energies and damping rates of
slow-moving quarks at next-to-leading order in the context of massless QCD
at high temperature using the real-time formulation of the fully-dressed
hard-thermal-loop perturbative expansion. These quantities are extracted
from the poles of the quark propagators. At lowest order, the energies $
\omega _{\pm }\left( p\right) $ are real, see Eq.~(\ref
{lowest-order-quark-energies}). The next-to-leading order contributions
necessitate the determination of the next-to-leading order quark
self-energies $\Sigma _{\pm }^{(1)}$, see Eq.~(\ref{NLO-complex-energy}).

In this work, we have given the analytic expression of $\Sigma _{\pm }^{(1)}$
in terms of loop-four-momentum integrals involving fully-HTL-dressed quark
and gluon propagators and vertex functions, see Eqs. (\ref{Sigma_1pm}) and (
\ref{Sigma_2pm}). These expressions were already written in \cite
{carrington--PRD75-2007-045019}. The HTL vertex functions themselves are\
derived and written as solid-angle integrals. We have rewritten these latter
using the Feynman parametrization in order to perform these solid-angle
integrals.

The next step is to perform the integrals. This is done numerically. The
usual approach is to use the spectral decomposition of the dressed
propagators. But we refrain from doing so and will try to tackle these
integrals head on. This is known to be difficult. One main difficulty is the
jumps the integrands suffer from at the singular lines of the propagators,
more prominently as the regularizer $\varepsilon \rightarrow 0$. We have
used a prototype integral to indicate how such difficulties might be
overcome. By splitting the integration region into appropriate domains, we
can obtain an estimate of the prototype integral with a robust behavior down
to $\epsilon \simeq 10^{-8}$ in unit of the quark thermal mass $m_{f}$.

Of course there are many other terms to handle, more involved than the
prototype. Will there be stability for each? Can we add them all? This is
currently under investigation.

\appendix

\section{Hard Thermal Loop Dressed Vertex Functions}

Discrepancies between different results in the literature \cite
{defu-et-al--PRD61-2000-085013,fueki-nakkagawa-yokota-yoshida} regarding the
derivation of the three and four-point hard-thermal-loop vertex functions in
the CTP formalism imposed on us an ab initio recalculation of these
quantities. We recover the results of \cite{fueki-nakkagawa-yokota-yoshida}.
This appendix summarizes our steps, with a notation close to \cite
{fueki-nakkagawa-yokota-yoshida}.

\subsection{Three-Point Vertex Functions}

The quark-gluon vertex functions are the sum of the bare vertices and the
corresponding hard thermal loops: 
\begin{equation}
\Gamma ^{\mu }=\gamma ^{\mu }+\delta \Gamma ^{\mu }.
\label{3-quark-gluon-vertex}
\end{equation}
In the \{12\} basis of the CTP formalism, the bare vertex $\gamma ^{\mu }$
is given by the relations: 
\begin{equation}
\gamma _{ijk}^{\mu }=\left\{ 
\begin{tabular}{l}
$\left( -1\right) ^{i-1}\gamma ^{\mu }\quad \mathrm{when}\quad i=j=k$ \\ 
$0\quad \mathrm{otherwise}$
\end{tabular}
\right. .  \label{3-bare-quark-gluon-vertex}
\end{equation}
The indices $i$, $j$, and $k$ take the values 1 and 2. The hard-thermal-loop
contributions $\delta \Gamma ^{\mu }$ to these vertices are obtained by
calculating the following one-loop diagrams: 
\begin{equation}
\delta \Gamma _{ijk}^{\mu }\left( P,Q,R\right) =4ig^{2}C_{F}\int \frac{
d^{4}K }{\left( 2\pi \right) ^{4}}K^{\mu }\NEG{K}V_{ijk}^{\prime
}-2ig^{2}N_{c}\int \frac{d^{4}K}{\left( 2\pi \right) ^{4}}K^{\mu }\NEG
{K}\left( V_{ijk}^{\prime }+V_{ijk}\right) ,  \label{3-quark-gluon-htl}
\end{equation}
in which the loop momentum $K$ is hard, in front of which external momenta
are neglected. In this expression, 
\begin{eqnarray}
V_{ijk} &=&\left( -1\right) ^{i+j+k-3}\bar{D}_{ij}\left( K\right)
D_{jk}\left( K-Q\right) D_{ki}\left( K+P\right) ;  \notag \\
V_{ijk}^{\prime } &=&\left( -1\right) ^{i+j+k-3}D_{ij}\left( K\right) \bar{D}
_{jk}\left( K-Q\right) \bar{D}_{ki}\left( K+P\right) .
\label{Vfunctions-in-3HTL-vertex}
\end{eqnarray}
The functions $D_{ij}\left( K\right) $ are the four \{12\} components of the
bare bosonic propagator: 
\begin{eqnarray}
D\left( K\right) &=& 
\begin{pmatrix}
D_{11}\left( K\right) & D_{12}\left( K\right) \\ 
D_{21}\left( K\right) & D_{22}\left( K\right)
\end{pmatrix}
\notag \\
&=& 
\begin{pmatrix}
\frac{1+n_{B}\left( k_{0}\right) }{K^{2}+i\varepsilon }-\frac{n_{B}\left(
k_{0}\right) }{K^{2}-i\varepsilon } & \frac{\theta \left( -k_{0}\right)
+n_{B}\left( k_{0}\right) }{K^{2}+i\varepsilon }-\frac{\theta \left(
-k_{0}\right) +n_{B}\left( k_{0}\right) }{K^{2}-i\varepsilon } \\ 
\frac{\theta \left( k_{0}\right) +n_{B}\left( k_{0}\right) }{
K^{2}+i\varepsilon }-\frac{\theta \left( k_{0}\right) +n_{B}\left(
k_{0}\right) }{K^{2}-i\varepsilon } & \frac{n_{B}\left( k_{0}\right) }{
K^{2}+i\varepsilon }-\frac{1+n_{B}\left( k_{0}\right) }{K^{2}-i\varepsilon }
\end{pmatrix}
.  \label{bosonic-bare-propagator}
\end{eqnarray}
The quantity $\bar{D}_{ij}\left( K\right) $ is $D_{ij}\left( K\right) $ with 
$n_{B}\left( k_{0}\right) $ replaced by $-n_{F}\left( k_{0}\right) $. We
write these quantities in terms of advanced ($A$), retarded ($R$) and
symmetric ($F$) propagators: 
\begin{eqnarray}
D_{11} &=&\frac{1}{2}\left( F+A+R\right) ;\quad D_{12}=\frac{1}{2}\left(
F+A-R\right) ;  \notag \\
D_{21} &=&\frac{1}{2}\left( F-A+R\right) ;\quad D_{22}=\frac{1}{2}\left(
F-A-R\right) ,  \label{Ds-in-terms-of-F-A-R}
\end{eqnarray}
where: 
\begin{eqnarray}
R\left( K\right) &=&\frac{\theta \left( k_{0}\right) }{K^{2}+i\varepsilon }+ 
\frac{\theta \left( -k_{0}\right) }{K^{2}-i\varepsilon };  \notag \\
A\left( K\right) &=&\frac{\theta \left( -k_{0}\right) }{K^{2}+i\varepsilon }
+\frac{\theta \left( k_{0}\right) }{K^{2}-i\varepsilon };  \notag \\
F\left( K\right) &=&\left( 1\pm 2n_{B,F}\left( k_{0}\right) \right) \left( 
\frac{1}{K^{2}+i\varepsilon }-\frac{1}{K^{2}-i\varepsilon }\right) .
\label{R-A-F-propagators}
\end{eqnarray}
The vertex functions in the \{$RA$\} basis are linearly related to the same
functions defined in the \{12\} basis. For example, the three-point vertex
function $\Gamma _{RAA}$ is given by the linear relation: 
\begin{equation}
\Gamma _{RAA}=\Gamma _{111}+\Gamma _{112}+\Gamma _{121}+\Gamma _{122}.
\label{Gamma_RAA}
\end{equation}
This relation applies to the kernels $V$ and $V^{\prime }$. Using the
expressions (\ref{Vfunctions-in-3HTL-vertex}) of the functions $V$ and $
V^{\prime }$, and the relations (\ref{Ds-in-terms-of-F-A-R}), we arrive at
the following expression: 
\begin{equation}
V_{RAA}=\frac{1}{2}
(A_{1}A_{2}A_{3}+R_{1}R_{2}R_{3}+F_{1}A_{2}A_{3}+R_{1}F_{2}A_{3}+R_{1}R_{2}F_{3}),
\label{V_RAA}
\end{equation}
where, for short, 1, 2 and 3 denote the arguments $K$, $K-Q$ and $K+P$
respectively. Next is to integrate over the internal momentum $K$. This is
performed in two steps: first over $k_{0}$, done using the residue theorem
in the $k_{0}$-complex plane. In this regard, the terms $A_{1}A_{2}A_{3}$
and $R_{1}R_{2}R_{3}$ give zero contribution each. The three others give the
following contributions: 
\begin{eqnarray}
FAA &=&\frac{+i}{8\pi ^{2}}\int_{0}^{+\infty }kdkN_{B,F}(k)\int \frac{
d\Omega _{s}}{4\pi }\frac{S^{\mu }S\hspace{-7pt}/}{\left( PS-i\varepsilon
\right) \left( QS+i\varepsilon \right) };  \notag \\
RFA &=&\frac{-i}{8\pi ^{2}}\int_{0}^{+\infty }kdkN_{B,F}(k)\int \frac{
d\Omega _{s}}{4\pi }\frac{S^{\mu }S\hspace{-7pt}/}{\left( QS+i\varepsilon
\right) \left( \left( P+Q\right) S-i\varepsilon \right) };  \notag \\
RRF &=&\frac{-i}{8\pi ^{2}}\int_{0}^{+\infty }kdkN_{B,F}(k)\int \frac{
d\Omega _{s}}{4\pi }\frac{S^{\mu }S\hspace{-7pt}/}{\left( PS-i\varepsilon
\right) \left( \left( P+Q\right) S-i\varepsilon \right) }.
\label{FAA-and-others}
\end{eqnarray}
Here, the quantities $N_{B,F}(k)$ are defined in Eq.~(\ref
{thermal-distributions}) and $S=\left( 1,\hat{s}\right) $ is a time-like
unit four-vector in which $\hat{s}$ is nothing but $\vec{k}/k$. Remember
that every external momentum is neglected in front of $K$. When summing the
contributions together, some will add up to a product of propagator
denominators multiplied by $i\varepsilon $, so are dropped, and some will
stand. In the end, we find: 
\begin{equation}
\delta \Gamma _{RAA}=\frac{-g^{2}C_{F}}{2\pi ^{2}}\int_{0}^{+\infty
}kdk\left( n_{B}\left( k\right) +n_{F}\left( k\right) \right) \int \frac{
d\Omega }{4\pi }\frac{S^{\mu }S\hspace{-7pt}/}{\left( PS-i\varepsilon
\right) \left( QS+i\varepsilon \right) }.  \label{htl_RAA}
\end{equation}
Now the integration over $k$ can be done. Using the known results: 
\begin{equation}
\int_{0}^{+\infty }dkkn_{B}\left( k\right) =\frac{\pi ^{2}}{6}T^{2};\quad
\int_{0}^{+\infty }dkkn_{F}\left( k\right) =\frac{\pi ^{2}}{12}T^{2},
\label{n_B-F_integrated}
\end{equation}
we finally find the expression for this hard thermal loop: 
\begin{equation}
\delta \Gamma _{RAA}\left( P,Q,R\right) =-m_{f}^{2}\int \frac{d\Omega }{4\pi 
}\frac{S^{\mu }S\hspace{-7pt}/}{\left( PS-i\varepsilon \right) \left(
QS+i\varepsilon \right) },  \label{htl_RAA-final}
\end{equation}
with $m_{f}=\sqrt{C_{F}/8}gT$. The other three-point HTL vertex functions
are obtained following a similar procedure. The relations between vertices
in the \{RA\} and \{12\} bases are: 
\begin{eqnarray}
\Gamma _{ARA} &=&\Gamma _{111}+\Gamma _{112}+\Gamma _{211}+\Gamma _{212}; 
\notag \\
\Gamma _{AAR} &=&\Gamma _{111}+\Gamma _{121}+\Gamma _{211}+\Gamma _{221}; 
\notag \\
\Gamma _{RRA} &=&\Gamma _{111}+\Gamma _{112}+\Gamma _{221}+\Gamma _{222}; 
\notag \\
\Gamma _{RAR} &=&\Gamma _{111}+\Gamma _{121}+\Gamma _{212}+\Gamma _{222}; 
\notag \\
\Gamma _{ARR} &=&\Gamma _{111}+\Gamma _{211}+\Gamma _{122}+\Gamma _{222}; 
\notag \\
\Gamma _{RRR} &=&\Gamma _{111}+\Gamma _{122}+\Gamma _{212}+\Gamma _{221},
\label{the-other-Gamma_RA}
\end{eqnarray}
with $\Gamma _{AAA}=0$ identically. For each of these vertex functions, one
works out steps similar to the ones for $\Gamma _{RAA}$ and one finds: 
\begin{eqnarray}
\delta \Gamma _{ARA}\left( P,Q,R\right) &=&-m_{f}^{2}\int \frac{d\Omega _{s} 
}{4\pi }\frac{S^{\mu }S\hspace{-7pt}/}{\left( PS+i\varepsilon \right) \left(
QS-i\varepsilon \right) };  \notag \\
\delta \Gamma _{AAR}\left( P,Q,R\right) &=&-m_{f}^{2}\int \frac{d\Omega _{s} 
}{4\pi }\frac{S^{\mu }S\hspace{-7pt}/}{\left( PS+i\varepsilon \right) \left(
QS+i\varepsilon \right) };  \notag \\
\delta \Gamma _{RRR}\left( P,Q,R\right) &=&-m_{f}^{2}\int \frac{d\Omega _{s} 
}{4\pi }\frac{S^{\mu }S\hspace{-7pt}/}{\left( PS-i\varepsilon \right) \left(
QS-i\varepsilon \right) };  \notag \\
\delta \Gamma _{RRA}\left( P,Q,R\right) &=&\delta \Gamma _{RAR}\left(
P,Q,R\right) =\delta \Gamma _{ARR}\left( P,Q,R\right) =0.
\label{the-other-HTL-3vertex}
\end{eqnarray}

\subsection{Four-Point Vertex Functions}

The two-gluon-two-quark vertex functions are all hard thermal loops, no bare
terms, given in the \{12\} basis by the relation\footnote{
A typo in (3.22) of \cite{fueki-nakkagawa-yokota-yoshida} is corrected.}: 
\begin{eqnarray}
\delta \Gamma _{ijkl}^{\mu \nu }\left( P,Q,R,U\right) &=&-8ig^{2}\left(
-1\right) ^{i+j+k+l}\int \frac{d^{4}K}{\left( 2\pi \right) ^{4}}K^{\mu
}K^{\nu }\NEG{K}  \notag \\
&&\times \left[ C_{F}D_{ij}\left( K\right) \bar{D}_{jk}\left( K-Q\right) 
\bar{D}_{kl}\left( K+P+U\right) \bar{D}_{li}\left( K+P\right) \right.  \notag
\\
&&-N_{c}\bar{D}_{ij}\left( K\right) D_{jk}\left( K-Q\right) D_{kl}\left(
K+P+U\right) D_{li}\left( K+P\right)  \notag \\
&&\left. +\frac{1}{2}N_{c}D_{li}(K-P)D_{jl}(K-P-U)\bar{D}_{kj}(K+R)\bar{D}
_{ik}(K)\right] .  \label{4-quark-gluon-vertex}
\end{eqnarray}
Consider for example the \{$RA$\}-component $\delta \Gamma _{RAAA}$, given
by: 
\begin{equation}
\delta \Gamma _{RAAA}=\delta \Gamma _{1111}+\delta \Gamma _{1112}+\delta
\Gamma _{1121}+\delta \Gamma _{1211}+\delta \Gamma _{1122}+\delta \Gamma
_{1212}+\delta \Gamma _{1221}+\delta \Gamma _{1222}.  \label{Gamma-RAAA}
\end{equation}
Consider the term multiplying $C_{F}$ in Eq.~(\ref{4-quark-gluon-vertex})
and call it $\delta \Gamma _{RAAA}^{1}$. Using the decompositions in Eq.~( 
\ref{Ds-in-terms-of-F-A-R}), we obtain, in a similar symbolic notation as in
Eq.~(\ref{V_RAA}), the relation: 
\begin{equation}
\delta \Gamma _{RAAA}^{1}=\frac{1}{2}F_{1}A_{2}A_{3}A_{4}+\frac{1}{2}
R_{1}F_{2}A_{3}A_{4}+\allowbreak \frac{1}{2}R_{1}R_{2}F_{3}A_{4}+\frac{1}{2}
R_{1}R_{2}R_{3}F_{4}.  \label{Gamma-RAAA-bis}
\end{equation}
The subscripts 1,2,3, and 4 stand for the momenta $K$, $K-Q$, $K+P+U$, and $
K+P$ respectively. The two contributions $\frac{1}{2}A_{1}A_{2}A_{3}A_{4}$
and $\frac{1}{2}R_{1}R_{2}R_{3}R_{4}$ to $\delta \Gamma _{RAAA}^{1}$ yield
zero each in the $k_{0}$-complex-plane integration and are not displayed
explicitly in Eq.~(\ref{Gamma-RAAA-bis}). The different terms in Eq.~(\ref
{Gamma-RAAA-bis}) contribute as follows: 
\begin{eqnarray}
FAAA &=&\int_{0}^{+\infty }\frac{kdk}{16\pi ^{2}}\int \frac{d\Omega _{s}} {
4\pi }\frac{iN_{B}(k)S^{\mu }S^{\nu }S\hspace{-7pt}/}{\left( \left(
2P+U\right) S+i\varepsilon \right) \left( \left( P-R\right) S+i\varepsilon
\right) \left( PS+i\varepsilon \right) };  \notag \\
RFAA &=&\int_{0}^{+\infty }\frac{kdk}{16\pi ^{2}}\int \frac{d\Omega _{s}}{
4\pi }\frac{-iN_{B}(k)K^{\mu }K^{\nu }S\hspace{-7pt}/}{\left( \left(
2P+U\right) S+i\varepsilon \right) \left( \left( R+P+U\right) S-i\varepsilon
\right) \left( \left( P+U\right) S-i\varepsilon \right) };  \notag \\
RRFA &=&\int_{0}^{+\infty }\frac{kdk}{16\pi ^{2}}\int \frac{d\Omega _{s}}{
4\pi }\frac{iN_{F}(k)S^{\mu }S^{\nu }S\hspace{-7pt}/}{\left( \left(
R-P\right) S-i\varepsilon \right) \left( \left( R+P+U\right) S-i\varepsilon
\right) \left( RS+i\varepsilon \right) };  \notag \\
RRRF &=&\int_{0}^{+\infty }\frac{kdk}{16\pi ^{2}}\int \frac{d\Omega _{s}}{
4\pi }\frac{iN_{F}(k)S^{\mu }S^{\nu }S\hspace{-7pt}/}{\left( PS+i\varepsilon
\right) \left( \left( P+U\right) S-i\varepsilon \right) \left(
RS+i\varepsilon \right) }.  \label{FAAA-and-others}
\end{eqnarray}
Adding these terms together yield the following result: 
\begin{equation}
\delta \Gamma _{RAAA}^{1}=\frac{i}{8\pi ^{2}}\int_{0}^{+\infty }kdk\int 
\frac{d\Omega _{s}}{4\pi }\frac{\left( n_{B}\left( k\right) +n_{F}\left(
k\right) \right) S^{\mu }S^{\nu }S\hspace{-7pt}/}{\left( PS-i\varepsilon
\right) \left( QS+i\varepsilon \right) \left( \left( P+U\right)
S-i\varepsilon \right) }.  \label{Gamma-RAAA-1}
\end{equation}
The integrations over $k$ can now be done using Eq.~(\ref{n_B-F_integrated}
). The terms multiplying $N_{c}$ and $N_{c}/2$ in Eq.~(\ref
{4-quark-gluon-vertex}) are worked out in a similar way; they cancel each
other. We therefore have the hard-thermal-loop four-vertex function: 
\begin{eqnarray}
\delta \Gamma _{RAAA}\left( P,Q,R,U\right) &=&m_{f}^{2}\int \frac{d\Omega
_{s}}{4\pi }\frac{S^{\mu }S^{\nu }S\hspace{-7pt}/}{\left( PS-i\varepsilon
\right) \left( QS+i\varepsilon \right) }  \notag \\
&&\times \left[ \frac{1}{\left( \left( P+U\right) S-i\varepsilon \right) }+ 
\frac{1}{\left( \left( P+R\right) S-i\varepsilon \right) }\right] .
\label{Gamma-RAAA-final}
\end{eqnarray}
The other \{$RA$\} four-vertex hard thermal loops are worked out in a
similar way. With $\delta \Gamma _{AAAA}=0$ and using the linear
relationships: 
\begin{eqnarray}
\delta \Gamma _{ARAA} &=&\delta \Gamma _{1111}+\delta \Gamma _{1112}+\delta
\Gamma _{1121}+\delta \Gamma _{2111}+\delta \Gamma _{1122}+\delta \Gamma
_{2112}+\delta \Gamma _{2121}+\delta \Gamma _{2122};  \notag \\
\delta \Gamma _{AARA} &=&\delta \Gamma _{1111}+\delta \Gamma _{1112}+\delta
\Gamma _{1211}+\delta \Gamma _{2111}+\delta \Gamma _{1212}+\delta \Gamma
_{2112}+\delta \Gamma _{2211}+\delta \Gamma _{2212};  \notag \\
\delta \Gamma _{AAAR} &=&\delta \Gamma _{1111}+\delta \Gamma _{1121}+\delta
\Gamma _{1211}+\delta \Gamma _{2111}+\delta \Gamma _{1221}+\delta \Gamma
_{2121}+\delta \Gamma _{2211}+\delta \Gamma _{2221};  \notag \\
\delta \Gamma _{RRAA} &=&\delta \Gamma _{1111}+\delta \Gamma _{1112}+\delta
\Gamma _{1121}+\delta \Gamma _{1122}+\delta \Gamma _{2211}+\delta \Gamma
_{2212}+\delta \Gamma _{2221}+\delta \Gamma _{2222};  \notag \\
\delta \Gamma _{RARA} &=&\delta \Gamma _{1111}+\delta \Gamma _{1112}+\delta
\Gamma _{1211}+\delta \Gamma _{1212}+\delta \Gamma _{2121}+\delta \Gamma
_{2122}+\delta \Gamma _{2221}+\delta \Gamma _{2222};  \notag \\
\delta \Gamma _{RAAR} &=&\delta \Gamma _{1111}+\delta \Gamma _{1121}+\delta
\Gamma _{1211}+\delta \Gamma _{1221}+\delta \Gamma _{2112}+\delta \Gamma
_{2122}+\delta \Gamma _{2212}+\delta \Gamma _{2222},
\label{the-other-4Gamma_RA}
\end{eqnarray}
one obtains the following results: 
\begin{eqnarray}
\delta \Gamma _{ARAA}\left( P,Q,R,U\right) &=&m_{f}^{2}\int \frac{d\Omega
_{s}}{4\pi }\frac{S^{\mu }S^{\nu }S\hspace{-7pt}/}{\left( PS+i\varepsilon
\right) \left( QS-i\varepsilon \right) }  \notag \\
&&\times \left[ \frac{1}{\left( \left( P+U\right) S+i\varepsilon \right) }+ 
\frac{1}{\left( \left( P+R\right) S+i\varepsilon \right) }\right] ;  \notag
\\
\delta \Gamma _{AARA}\left( P,Q,R,U\right) &=&m_{f}^{2}\int \frac{d\Omega
_{s}}{4\pi }\frac{S^{\mu }S^{\nu }S\hspace{-7pt}/}{\left( PS+i\varepsilon
\right) \left( QS+i\varepsilon \right) }  \notag \\
&&\times \left[ \frac{1}{\left( \left( P+U\right) S+i\varepsilon \right) }+ 
\frac{1}{\left( \left( P+R\right) S-i\varepsilon \right) }\right] ;  \notag
\\
\delta \Gamma _{AAAR}\left( P,Q,R,U\right) &=&m_{f}^{2}\int \frac{d\Omega
_{s}}{4\pi }\frac{S^{\mu }S^{\nu }S\hspace{-7pt}/}{\left( PS+i\varepsilon
\right) \left( QS+i\varepsilon \right) }  \notag \\
&&\times \left[ \frac{1}{\left( \left( P+U\right) S-i\varepsilon \right) }+ 
\frac{1}{\left( \left( P+R\right) S+i\varepsilon \right) }\right] ;  \notag
\\
\delta \Gamma _{RRAA}\left( P,Q,R,U\right) &=&\delta \Gamma _{RARA}\left(
P,Q,R,U\right) =\delta \Gamma _{RAAR}\left( P,Q,R,U\right) =0.
\label{the-other-htl-4vertex}
\end{eqnarray}

The remaining eight components can be either calculated directly or
obtained\ from those above using the KMS conditions.

\subsection{Change of Notation}

Finally, as we mentioned early in this appendix, the notation we use here
for the vertex functions is close to that used in \cite
{fueki-nakkagawa-yokota-yoshida}. However, the notation we use in the main
text is close to the one used in \cite{carrington-et-al--PRD78-2008-045018}.
They are related in the following manner: 
\begin{eqnarray}
\Gamma _{I_{1}I_{2}I_{3}}^{\mu }\left( P_{1},P_{2},P_{3}\right) &=&\Gamma
_{i_{1}i_{3}i_{2}}^{\mu }\left( P_{1},-P_{2}\right) ;  \notag \\
\Gamma _{I_{1}I_{2}I_{3}I_{4}}^{\mu \nu }\left(
P_{1},P_{2},P_{3},P_{4}\right) &=&\Gamma _{i_{1}i_{4}i_{3}i_{2}}^{\mu \nu
}\left( P_{1},P_{4},P_{3},-P_{2}\right) ,  \label{notation-conversion}
\end{eqnarray}
with the understanding $I_{j}=R(A)\leftrightarrow i_{j}=\mathrm{a}(\mathrm{r}
)$. Thus, for the three-vertex functions: 
\begin{eqnarray}
\Gamma _{\mathrm{arr}}^{\mu }\left( P,Q\right) &=&\Gamma _{RAA}^{\mu }\left(
P,-Q,R\right) =\gamma ^{\mu }+I_{--}^{\mu }\left( P,Q\right) ;  \notag \\
\Gamma _{\mathrm{rar}}^{\mu }\left( P,Q\right) &=&\Gamma _{AAR}^{\mu }\left(
P,-Q,R\right) =\gamma ^{\mu }+I_{+-}^{\mu }\left( P,Q\right) ;  \notag \\
\Gamma _{\mathrm{aar}}^{\mu }\left( P,Q\right) &=&\Gamma _{RAR}^{\mu }\left(
P,-Q,R\right) =0;  \notag \\
\Gamma _{\mathrm{rra}}^{\mu }\left( P,Q\right) &=&\Gamma _{ARA}^{\mu }\left(
P,-Q,R\right) =\gamma ^{\mu }+I_{++}^{\mu }\left( P,Q\right) ;  \notag \\
\Gamma _{\mathrm{ara}}^{\mu }\left( P,Q\right) &=&\Gamma _{RRA}^{\mu }\left(
P,-Q,R\right) =0,  \label{3-vertices-ar-RA}
\end{eqnarray}
with the following definition: 
\begin{equation}
I_{\eta _{1}\eta _{2}}^{\mu }\left( P,Q\right) =m_{f}^{2}\int \frac{d\Omega
_{s}}{4\pi }\frac{S^{\mu }S\hspace{-7pt}/}{\left( PS+i\eta _{1}\varepsilon
\right) \left( QS+i\eta _{2}\varepsilon \right) }.  \label{I-mu-def}
\end{equation}
For the four-vertex functions we need in the text, we have: 
\begin{eqnarray}
\Gamma _{\mathrm{arrr}}^{\mu \nu }\left( P,K\right) &\equiv &\Gamma _{ 
\mathrm{arrr}}^{\mu \nu }\left( P,K,-K,P\right) =\Gamma _{RAAA}^{\mu \nu
}\left( P,-P,-Q,Q\right) =I_{--}^{\mu \nu }\left( P,K\right) ;  \notag \\
\Gamma _{\mathrm{aarr}}^{\mu \nu }\left( P,K\right) &\equiv &\Gamma _{ 
\mathrm{aarr}}^{\mu \nu }\left( P,K,-K,P\right) =\Gamma _{RAAR}^{\mu \nu
}\left( P,-P,-Q,Q\right) =0;  \notag \\
\Gamma _{\mathrm{arar}}^{\mu \nu }\left( P,K\right) &\equiv &\Gamma _{ 
\mathrm{arar}}^{\mu \nu }\left( P,K,-K,P\right) =\Gamma _{RARA}^{\mu \nu
}\left( P,-P,-Q,Q\right) =0,  \label{4vertex-we-need}
\end{eqnarray}
with the definition: 
\begin{eqnarray}
I_{\eta _{1}\eta _{2}}^{\mu \nu }\left( P,K\right) &=&m_{f}^{2}\int \frac{
d\Omega _{s}}{4\pi }\frac{-S^{\mu }S^{\nu }S\hspace{-7pt}/}{\left[ PS+i\eta
_{1}\epsilon \right] \left[ PS+i\eta _{2}\epsilon \right] }  \notag \\
&&\times \left[ \frac{1}{\left( P+K\right) S+i\eta _{1}\epsilon }+\frac{1}{
\left( P-K\right) S+i\eta _{2}\epsilon }\right] .  \label{I-mu-nu-def}
\end{eqnarray}
Note that both $I^{\mu }$ and $I^{\mu \nu }$ are already used in the main
text, see Eq.~(\ref{the-Is-definition}).


\begin{thebibliography}{99}
\bibitem{rhic-collabos} Yi Guo for the STAR collaboration,
J.~Phys.~Conf.~Ser.~535 (2014) 012006 \texttt{[arXiv:1407.6788 [hep-ex]]};
J. Adams \textit{et al}. [STAR Collaboration], Nucl.~Phys.~\textbf{A}757
(2005) 102 \texttt{[nucl-ex/0501009]}; K. Adcox \textit{et al}. [PHENIX
Collaboration], Nucl.~Phys.~\textbf{A}757 (2005) 184 \texttt{[nucl-ex/0410003]};
I. Arsene \textit{et al}. [BRAHMS Collaboration], Nucl.~
Phys.~\textbf{A}757 (2005) 1 \texttt{[nucl-ex/0410020]}; B.B. Back
\textit{et al}. [PHOBOS Collaboration], Nucl.~Phys.~\textbf{A}757 (2005) 28
\texttt{[nucl-ex/0410022]}.

\bibitem{alice} See J.F. Grosse-Oetringhaus (for the ALICE\ collaboration), 
\texttt{arXiv:1408.0414}, proceedings of Quark Matter 2014, for a recent
overview and references therein.

\bibitem{lqcd} H.T. Ding, \texttt{arXiv:1404.5134 [hep-lat]}; G. Aarts 
\textit{et al}., \texttt{arXiv:1403.5183 [hep-lat]}; C. Allton \textit{et al}.,
J.~Phys.~Conf.~Ser.~509 (2014) 012015 \texttt{[arXiv:1310.5135 [hep-lat]]};
S. Borsanyi {\it et al.}, Phys.~Lett.~{bf B} 370 (2014) 99 {\tt [arXiv:1309.5258 [hep-lat]]};
L. Levkova and C. DeTar, Phys.~Rev.~Lett.~112 (2014) 012002
\texttt{[arXiv:1309.1142 [hep-lat]]}; A. Amato \textit{et al}., Phys.~Rev.~Lett.~111
(2013) 172001 \texttt{[arXiv:1307.6763 [hep-lat]]}; S. Borsanyi \textit{et
al }., JHEP 1208 (2012) 126 \texttt{[arXiv:1205.0440 [hep-lat]]}; S.
Borsanyi \textit{et al}., JHEP 1208 (2012) 053 \texttt{[arXiv:1204.6710
[hep-lat]]}; S. Gupta \textit{et al}., Science 332 (2011) 1525 \texttt{
[arXiv:1105.3934 [hep-ph]]}; H.B. Meyer, Eur. Phys. J. \textbf{A}47 (2011)
86 \texttt{\ [arXiv:1104.3708 [hep-lat]]}; S. Borsanyi \textit{et al}., JHEP
1011 (2010) 077 \texttt{[arXiv:1007.2580 [hep-lat]]};
Y. Aoki {\it et al.}, Phys.~Lett.~{\bf B}643 (2006) 46 {\tt [hep-lat/0609068]};
Y. Aoki {\it et al.}, Nature 443 (2006) 675 {\tt [hep-lat/0611014]};
Z. Fodor and S. Katz, JHEP 0404 (2004) 050 \texttt{[hep-lat/0402006]}.

\bibitem{hydro} M. Elias, J. Peralta-Ramos, E. Calzetta Phys. Rev. \textbf{D}
90 (2014)\ 014038 \texttt{[arXiv:1404.7790 [hep-ph]]}; H. Song, \texttt{\
arXiv:1401.0079 [nucl-th]}; E. Calzetta, J. Peralta-Ramos Phys.~Rev.~\textbf{
\ D}88 (2013) 095010 \texttt{[arXiv:1309.5412 [hep-ph]]}; L. Del Zanna 
\textit{\ et al}., Eur.~Phys.~J. \textbf{C}73 (2013) 2524 \texttt{
[arXiv:1305.7052 [nucl-th]]}; J. Peralta-Ramos, E. Calzetta, Phys.~Rev.~ 
\textbf{D}86 (2012) 125024 \texttt{[arXiv:1208.2715 [hep-ph]]}; J.
Peralta-Ramos, E. Calzetta, Eur.~Phys.~J. \textbf{A}48 (2012) 163 \texttt{
[arXiv:1207.2396 [nucl-th]]}; Phys.~Rev.~\textbf{C}82 (2010) 054905 \texttt{
[arXiv:1003.1091 [hep-ph]]}; \texttt{arXiv:0908.3656 [nucl-th]}; A. K.
Chaudhuri, J.~Phys.~\textbf{G}39 (2012) 125102 \texttt{[arXiv:1111.5713
[nucl-th]]}; Phys.~Rev.~\textbf{C}82 (2010) 047901 \texttt{[arXiv:1006.4478
[nucl-th]]}; C. Shen, U. Heinz, P. Huovinen, and H. Song, Phys.~Rev.~\textbf{
C}84 (2011) 044903 \texttt{\ [arXiv:1105.3226 [nucl-th]]}; Y. Akamatsu, T.
Hatsuda, and T. Hirano, Nucl.~Phys.~\textbf{A}830 (2009) 865 \texttt{
[arXiv:0907.2981 [hep-ph]]}; U. Heinz, \texttt{nucl-th/0512051}.

\bibitem{HTL-perturbation} N. Su, Commun.~Theor.~Phys.~57 (2012) 409
\texttt{[arXiv:1204.0260 [hep-ph]]}; J.O. Andersen and M. Strickland, Ann.~Phys.~
317 (2005) 281 \texttt{[hep-ph/0404164]}; U. Kraemmer and A. Rebhan, Rep.~
Prog.~Phys.~\textbf{67} (2004) 351 \texttt{[hep-ph/0310337]}; M. Le Bellac,
` \textit{Thermal Field Theory}', Cambridge Univ.~Press,~1996; E. Braaten
and R. D. Pisarski, Nucl.~Phys.~\textbf{B}339 (1990) 310, Nucl.~Phys.~ 
\textbf{B} 337 (1990) 569, Phys.~Rev.~Lett.~\textbf{64} (1990) 1338; J.
Frenkel and J. C. Taylor, Nucl.~Phys.~\textbf{B}334 (1990) 199.

\bibitem{two-three-loop-htl} N. Haque \textit{et al}., JHEP 1405 (2014) 027 
\texttt{[arXiv:1402.6907 [hep-ph]]}; N. Haque \textit{et al}., Phys.~Rev.~ 
\textbf{D}89 (2014) 061701(R) \texttt{[arXiv:1309.3968 [hep-ph]]}; N. Haque,
M.G. Mustafa, and M. Strickland, JHEP 1307 (2013) 184
\texttt{[arXiv:1302.3228 [hep-ph]]}; Phys.~Rev.~\textbf{D}87 (2013) 105007
\texttt{[arXiv:1212.1797 [hep-ph]]}; J.O. Andersen, L.E. Leganger, M. Strickland,
and N. Su, JHEP 1108 (2011) 053 \texttt{[arXiv:1103.2528 [hep-ph]]}; M.
Strickland, J. O. Andersen, L. E. Leganger, and N. Su, Prog.~Theor.~Phys.~
Suppl.~187 (2011) 106 \texttt{[arXiv:1011.0416 [hep-ph]]}; J. O. Andersen,
L.E. Leganger, M. Strickland, and N. Su, Phys.~Lett.~\textbf{B}696 (2011)
468 \texttt{[arXiv:1009.4644 [hep-ph]]}; Y. Jiang, H.X. Zhu, W.M. Sun, and
H.S. Zong, J.~Phys.~\textbf{G}37 (2010) 055001 \texttt{[arXiv:1003.5031 [hep-ph]]}.

\bibitem{liu-luo-wang-xu} J. Liu, M.J. Luo, Q. Wang and H.J. Xu, Phys.~Rev.~ 
\textbf{D}84 (2011) 125027 \texttt{[arXiv:1109.4083 [hep-ph]]}.

\bibitem{early-htl} E. Braaten and R. D. Pisarski, J. Frenkel and J. C.
Taylor, cited in \cite{HTL-perturbation}.

\bibitem{gamt0} E. Braaten and R.D. Pisarski, Phys.~Rev.~\textbf{D}42 (1990)
R2156.

\bibitem{gamq0} R. Kobes, G. Kunstatter and K. Mak, Phys.~Rev.~\textbf{D}45
(1992) 4632; E. Braaten and R. Pisarski, Phys.~Rev.~\textbf{D}46 (1992) 1829.

\bibitem{carrington--PRD75-2007-045019} M.E. Carrington, Phys.~Rev.~
\textbf{D}75 (2007) 045019 [\texttt{arXiv:hep-ph/0610372}].

\bibitem{imaginary-time-formalism} J. I. Kapusta and C. Gale, `\textit{\
Finite-Temperature Field Theory}:\textit{\ Principle and Applications'},
Cambridge Monographs on Mathematical Physics, 2nd ed., 2011; M. LeBellac,
cited in \cite{HTL-perturbation}.

\bibitem{gaml} A. Abada. and O. Azi, Phys.~Lett.~\textbf{B}463 (1999) 117 
\texttt{[hep-ph/9807439]}; A. Abada, O. Azi and K. Benchallal, Phys.~Lett.~ 
\textbf{B}425 (1998) 158 \texttt{[hep-ph/9712210]}.

\bibitem{gamt} A. Abada, K. Bouakaz and O. Azi, Phys.~ Scri.~\textbf{74}
(2006) 77 \texttt{[hep-ph/0402041]}.

\bibitem{gamq} A. Abada, K. Bouakaz and N. Daira-Aifa, Int.~Jour.~Mod.~Phys. 
\textbf{A} 22 (2007) 6033; A. Abada, N. Daira-Aifa and K. Bouakaz, Int.~
Jour.~Mod.~Phys.~\textbf{A}21 (2006) 5317 \texttt{[hep-ph/0511258]}; A.
Abada, K. Bouakaz and N. Daira-Aifa, Eur.~Phys.~J.~\textbf{C}18 (2001) 765 
\texttt{[hep-ph/0008335]}.

\bibitem{fermions} K. Bouakaz and A. Abada, AIP Conf.~Proc.~1006 (2008) 150;
A. Abada, K. Bouakaz and D. Deghiche, Mod.~Phys.~Lett.~\textbf{A}22 (2007)
903; A. Abada and K. Bouakaz, \texttt{hep-ph/0209246}.

\bibitem{photons} A. Abada and N. Daira-Aifa, JHEP 04 (2012) 071\texttt{\
[arXiv:1112.6065[hep-ph]]}.

\bibitem{sqed} A. Abada and K. Bouakaz, JHEP \textbf{01} (2006) 161 \texttt{
\ [hep-ph/0510330]}.

\bibitem{schulz} H. Schulz, Nucl.~Phys.~\textbf{B}413 (1994) 353\texttt{\
[hep-ph/9306298]}.

\bibitem{carrington-et-al--EPJC50-2007-711} M.E. Carrington, T. Fugleberg,
D.S. Irvine, and D. Pickering, Eur.~Phys.~J.~C 50 (2007) 711 \texttt{\
[arXiv:hep-ph/0608298]}.

\bibitem{carrington-et-al--PRD78-2008-045018} M.E. Carrington, A. Gynther
and D. Pickering, Phys.~Rev.~\textbf{D}78 (2008) 045018
\texttt{[arXiv:0805.0170[hep-ph]]}.

\bibitem{real-time-formalism} M. LeBellac, in \cite{HTL-perturbation}; N.P.
Landsman and C.G. van Weert, Phys.~Rep.~145 (1987) 141; K.C. Chou, Z.B. Su,
B.L. Hao, and L. Yu, Phys.~Rep.~118 (1985) 1.

\bibitem{mirza-carrington--PRD87-2013-065008} A. Mirza and M.E. Carrington,
Phys.~Rev.~\textbf{D}87 (2013) 065008 \texttt{[arXiv:1302.3796[hep-ph]]}.

\bibitem{martin-schwinger} P.C. Martin, J. Schwinger, Phys.~Rev.~115 (1959)
1432.

\bibitem{keldysh} L.V. Keldysh, Sov.~Phys.~JETP 20 (1965) 1018.

\bibitem{defu-et-al--PRD61-2000-085013} H. Defu, M.E. Carrington, R. kobes,
and U. Heinz, Phys.~Rev.~\textbf{D}61 (2000) 085013
\texttt{[arXiv:hep-ph/9911494]}; Phys.~Rev.~\textbf{D}67 (2003) 049902.

\bibitem{fueki-nakkagawa-yokota-yoshida} Y. Fueki, H. Nakkagawa, H. Yokota,
K. Yoshida, Prog.~Theor.~Phys.~107 (2002) 759 \texttt{[arXiv:hep-ph/0111275]}.

\bibitem{axodraw} The Feynman diagrams are drawn using \texttt{jaxodraw}, D.
Binosi and L. Theussl, Comput.~Phys.~Commun.~161 (2004) 76 \texttt{
[arXiv:hep-ph/0309015]}.

\bibitem{defu-heinz--EPJC7-1999-101} H. Defu and U. Heinz, Eur.~Phys.~Jour.~ 
\textbf{C}7 (1999) 101 \texttt{[arXiv:hep-th/9710090]}.
\end{thebibliography}
\end{document}